\documentclass[a4paper,onecolumn,accepted=2020-08-20,11pt]{quantumarticle}
\pdfoutput=1

 \usepackage[pdfstartview=FitH]{hyperref}

\usepackage[numbers,sort&compress]{natbib}
\usepackage[utf8]{inputenc}
\usepackage[english]{babel}
\usepackage[T1]{fontenc}

\usepackage{amsmath,amssymb,amsthm}
\usepackage{graphicx}
\usepackage{braket}
\usepackage{empheq}
\usepackage{enumitem}
\usepackage{subfigure}
\usepackage{comment}
\usepackage{upgreek}
\usepackage[normalem]{ulem}
\usepackage{epstopdf}

\usepackage{doi}
\usepackage{wrapfig}
\usepackage{tikz}
\usepackage{lipsum}
\usepackage{dsfont}

\begin{document}

\title{Completely Positive, Simple, and Possibly
Highly Accurate Approximation of the Redfield Equation}

\author{Dragomir Davidovi\'c}
\affiliation{Georgia Institute of Technology, Atlanta, Georgia, United States}
\orcid{0000-0003-1486-4910}
\email{dragomir.davidovic@physics.gatech.edu}
\maketitle

\begin{abstract}

Here we present a Lindblad master equation that approximates the Redfield equation, a well known master equation derived from first principles, without
significantly compromising the range of applicability of the Redfield equation.  Instead of full-scale coarse-graining, this approximation only truncates terms
in the Redfield equation that average out over a time-scale typical of the quantum system.
The first step in this approximation is to properly renormalize the system Hamiltonian,
to symmetrize the gains and losses of the state due to the environmental coupling.
In the second step, we swap out an arithmetic mean of the spectral density with a geometric one, in these gains and losses,
thereby restoring complete positivity. This completely positive approximation, GAME (geometric-arithmetic master equation), is adaptable between its time-independent,
time-dependent, and Floquet form.
In the exactly solvable, three-level, Jaynes-Cummings model, we find that the error of the approximate state is
almost an order of magnitude lower than that obtained by solving the coarse-grained stochastic master equation.
As a test-bed, we use a ferromagnetic Heisenberg spin-chain with long-range dipole-dipole coupling between up to 25-spins, and study the differences between various master equations. We find that GAME
has the highest accuracy per computational resource.

\end{abstract}
\maketitle

\section{\label{sec:intro}Introduction}

Quantum correlations between particles in quantum systems are counterintuitive, giving us exotic algorithms that can greatly
accelerate some calculations using quantum
computing as opposed to conventional computing.
It remains an unsettled issue, however, to describe correlated quantum dynamics in large many-body quantum systems,
with the precision required for quantum computing, if the systems are coupled to an environment.
In the Born-Markov approximation, the time
dependence of the reduced state of the system can be approximated by a first-order differential equation, (e.g., the master equation),
which greatly simplifies this description.~\cite{Gardiner,BreuerHeinz-Peter1961-2007TToO}
The Born-Markov approximation is justified
if the time in which the environment changes the system state in the interaction picture, $\tau_r$, is much longer than the time at which the correlations in the environment decay, $\tau_c$.~\cite{BreuerHeinz-Peter1961-2007TToO}
We will use the terms bath and environment interchangeably throughout this paper.

The master equation does not necessarily preserve the positivity of the reduced quantum state of the system.
There is only one form of the master equation that is a completely positive
map, the deeply respected Gorini-Kossakowski-Sudarshan-Lindblad (GKSL) equation.~\cite{lindblad1976,Gorini,BreuerHeinz-Peter1961-2007TToO}
The GKSL equation is a consequence of the
Dirac–von Neumann axioms, while the physical scope of the equation rests in the coarse-graining of the reduced system dynamics over a timescale much longer than $\tau_c$.~\cite{LIDAR200135}

It has been very difficult to derive the
GKSL equation from first principle calculations
using the total
Hamiltonian of the system and environment as a starting point.~\cite{Kohen}
The operator-projection technique has been the standard way to derive quantum master equations~\cite{Zwanzig,Nakajima,kubo,BreuerHeinz-Peter1961-2007TToO}
initially pioneered by Redfield.~\cite{Redfield,REDFIELD19651}
The Redfield equation, however, suffers from a lack of positivity preservation leading to negative
probabilities of observables~\cite{Kohen} which violates the mathematical definition of probability. Negative states also pose problems in terms of how to quantify the entanglement of the system, which is important for quantum computing.
One measure of entanglement is the negativity of the state's partial transpose,~\cite{vidal}
which raises the question: Can we trust negativity of the partial transpose as a genuine measure of entanglement if the approximate state itself is negative?

Although it has this problem of producing negative probabilities, the Redfield equation has major advantages as well. It is a highly accurate master equation that is  simple to apply
and not axiomatic. Thus, the Redfield equation and its modifications have found significant use in quantum chemistry, condensed matter physics, and quantum optics.~\cite{redfieldreview,Kohen,Ivan,Dassia,SCHRODER,Andres,Timm,Jeske,Bricker}
The issue of negative states and their relevance remains debated.~\cite{Whitney,Hartmann}
The Redfield equation with time dependent coefficients (TDC) has reduced negativity.
The nonpositivity of that equation can even be seen as a benefit of a red-flag for the breakdown of the master equation concept.~\cite{Hartmann} This benefit does not extend well to the Redfield equation with asymptotic coefficients as we will discuss in this paper. The latter is the widely used master equation that we are approximating into a GKSL form in this paper.

Our main focus here is to find a simple, completely positive approximation of the Redfield equation without significantly compromising its accuracy. Note that in the same weak coupling regime where the Redfield equation is valid,
quantum trajectory description is available which guarantees
positivity.~\cite{ting,Vega} There are also quite a few non-perturbative approaches such as the exact quasi adiabatic path integral method (QUAPI),~\cite{Makri,THORWART2004333,Nalbach} the hierarchical equations of motion (HEOM),~\cite{Tanimura,Tanimura1,Tanimura2,ZhenHua,Cheng_2015}, the
multiconfiguration time-dependent Hartree (MCTDH)~\cite{MEYER199073,BECK20001}, and the multilayer formulation ML-MCTDH.~\cite{Haobin,Jie} These are powerful approaches but can be very challenging. Exact methods aiming at improved efficiency include quantum trajectory based hierarchy of stochastic pure states (HOPS)~\cite{Suess,Pan-Pan,Hartmann1} and tensor network methods.~\cite{Strathearn,Florian} Other strong coupling approaches involve unitary transformations that result in a non-perturbative bath-renormalization of the system Hamiltonian, followed by a quantum master equation in the weak coupling limit after the renormalization.~\cite{Jang, McCutcheon}

There have been several attempts to cure the issue of negative states by
modifying the Redfield master equation into a GKSL form.
This was done for the first time by Davies,~\cite{davies1974} who gave us the
rotating-wave approximation (RWA), also known as the secular approximation. The
RWA amounts to coarse-graining of the reduced state on a timescale comparable to the Heisenberg time.
Since this time scales exponentially with system size,
the RWA does not work well for large many-body quantum systems
that generally have exponentially suppressed level spacings with increasing size of the system.
More recently, coarse-grained GKSL master equations have been derived from first principles~\cite{Majenz,Giovannetti,mozgunov} capable of capturing correlation effects that the RWA cannot.
Another notable completely positive approximation is the partial-RWA (PRWA).~\cite{Vogt,Tscherbul} An alternative approach to
restoring complete positivity includes the dynamically coarse-grained (DCG) master equation.~\cite{Schaller,Benatti}
Phenomenological models have also worked well to briskly set up a completely positive master equation.~\cite{Munro,Wilkie,Benoit,perlind}

The approximate, completely positive GAME works in two steps. First, it identifies an implicit renormalization of the Hamiltonian (e.g., the Lamb-shift) in the Redfield equation.
After separating the Lamb-shift by adding it to the system Hamiltonian, the second step separates the dissipator into two parts: a completely positive dissipator and the remainder.
The latter has the property that it averages-out on a short time-scale compared to the system relaxation time, and is therefore dropped,
while the renormalized Hamiltonian and the completely positive dissipator remain intact.

Practically, in terms of the spectral density (SD), we swap out an arithmetic mean of the SD at two system frequencies, with a geometric mean of the SD. The difference between the two means has the sought after property that it averages out on a typical time scale of the system.
We find that the state error added by the approximation is linear in system-bath coupling and generally comparable to the accuracy of the Redfield equation (when applicable).

GAME belongs to a relatively new group of approximations that we refer to here as the  $\sqrt{\text{SD}}$-approximations.\\ \cite{perlind,Nathan} In this paper we compare many of the aforementioned GKSL-equations with GAME, by applying them on a spin-1/2 Heisenberg chain, and have not found a single one that outperforms GAME in terms of the closeness of its solutions to those of the Redfield equation.

The renormalized Hamiltonian plays the key role and is the primary reason that GAME is capable of accounting for the
correlation effects that the bath introduces into the system dynamics. The effects of the renormalization are studied in detail in the case of the 3-level
Jaynes-Cummings model, where the eigenstates of the renormalized Hamiltonian are identified as the relaxation modes of the system coupled to the heat bath, those that decay with a single relaxation time.

By construction, GAME retains the simplicity of the Redfield equation, which is one of its assets.
Even more important is its combination of both simplicity and accuracy.

The  paper is organized as follows. In Sec.~\ref{secMES} we introduce the notation relevant for this work.
In Sec.~\ref{sec:RMA},~\ref{sec:CRE}, and~\ref{sec:RWA} we review the Redfield, Coarse-Grained Redfield, and the Davies-Lindblad master equations,
respectively, as a prelude to GAME, which is presented in Sec.~\ref{sec:game}.
In Sec.~\ref{sec:3levsys} we implement GAME on an exactly solvable
3-level Jaynes–Cummings model.
We follow by the detailed investigation of a 25-body spin-1/2 ferromagnetic spin-chain in Sec.~\ref{sec:spinchain}. We end the analysis by presenting a detailed  comparison with other GKSL equations in Sec.~\ref{sec:benchmark}, followed by discussion and conclusion.

\section{\label{secMES} Introducing the problem}

Perturbative master equations have been derived many times in the literature. For a tutorial and some
recent examples we refer the reader to Refs.~\cite{kryszewski2008master,Whitney,mozgunov,Hartmann}.

Here we consider a quantum system coupled to a heat-bath, represented by the total Hamiltonian
\begin{equation}
H_{tot}=H_0\otimes \mathds{1}_B+\mathds{1}_S \otimes H_B+A\otimes B.
\label{HamiltonianH}
\end{equation}
$H_0$ and $A$ are the respective Hamiltonian and arbitrary operators performing in the Hilbert space of the quantum system, $H_B$ and $B$ are the corresponding operators for the heat bath, and $A\otimes B$ is the
interaction Hamiltonian between the system and the heat bath. All the operators are assumed to be Hermitian. For notational clarity we only consider one coupling term in the interaction Hamiltonian. The results can be straightforwardly extended to a sum of interaction terms assuming uncorrelated heat baths.

In the Hilbert space of the quantum system, we use the eigenbasis of $H_0$, $H_0\vert n\rangle=E_n\vert n\rangle$, where $\vert n\rangle$ are the eigenvectors and $E_n$ are the eigenenergies.
Then, in the interaction picture $A(t)=e^{iH_0t}Ae^{-iH_0t}$ is decomposed as
\begin{equation}
A(t)=\sum_{n,m} A_{nm}e^{-i\omega_{mn}t}\vert n\rangle\langle m\vert,
\label{Adecomposition}
\end{equation}
where $A_{nm}=\langle n\vert A\vert m\rangle$, $\omega_{mn}=E_m-E_n$, $t$ is time, and we use $\hbar=1$.
Operator $A(t)$ can be represented in the matrix form as
\begin{equation}
A(t)=A\circ\Omega(t),
\label{Adecomposition1}
\end{equation}
where $\circ$ is the Hadamard product (the element-wise product) and $[\Omega(t)]_{nm}=e^{-i\omega_{mn}t}$.

The bath correlation function is defined as $C(t)=Tr[\rho_B B(t)B(0)]$, where $B(t)=e^{iH_Bt}Be^{-iH_Bt}$ and $\rho_B$ is the reduced density matrix of the heat bath. We work under the premise
that the excitations in the heat-bath decay on a time-scale much smaller than the time scale of the system state in the interaction picture. In that case, the change in the reduced density matrix of the bath due to the coupling to the system can be neglected, in the leading order of the approximation, and the density matrix of the environment can be approximated
by the thermal state where $C(t)=C^{\star}(-t)$.~\cite{BreuerHeinz-Peter1961-2007TToO}
In our notation we reserve the symbols $\rho$ and $\varrho$, for the density matrix in the Schr\"{o}dinger and interaction picture, respectively.

\begin{table}
\centering

\begin{tabular}{|c|c|c|c|}
  \hline
  $C(t)$ & $\gamma(\omega)$ & $S(\omega)$\\
  \hline
$ \frac{1}{2\pi} \int_{-\infty}^{\infty}e^{-i\Omega t} \gamma(\Omega)d\Omega$ & $\int_{-\infty}^{\infty}e^{i\omega t} C(t)dt$ & $\frac{1}{2\pi}\mathcal{P}\int_{-\infty}^{\infty}\frac{\gamma(\Omega)}{\omega-\Omega}d\Omega$\\
  \hline
  $ \frac{g\omega_c^2}{(1+i\omega_ct)^2}$ & $2\pi g\omega e^{-\frac{\omega}{\omega_c}}\Theta(\frac{\omega}{\omega_c})$ & $-g\omega_c\Big[1-\frac{\omega}{\omega_c} e^{-\frac{\omega}{\omega_c}}Ei(\frac{\omega}{\omega_c})\Big]$ \\

 \hline
$
   \begin{array}{l}
     -g\omega_c^2[\ln \vert\omega_c t\vert+\gamma\\+i\frac{\pi}{2}\text{sgn}(t)]+O[t^2\ln(t)]
   \end{array}
$
& $2\pi g\omega \frac{\omega_c^2}{\omega_c^2+\omega^2}\Theta(\frac{\omega}{\omega_c})$ & $-g\omega_c\Big[\frac{\pi/2-(\omega/\omega_c)\ln(|\omega|/\omega_c)}{1+(\omega/\omega_c)^2}\Big]$ \\
  \hline
  $ \frac{6g\omega_c^2}{(1+i\omega_ct)^4}$ & $2\pi g\frac{\omega^3}{\omega_c^2}e^{-\frac{\omega}{\omega_c}}\Theta(\frac{\omega}{\omega_c})$ & $\begin{array}{c}
-g\omega_c\Big[2+\frac{\omega}{\omega_c}+(\frac{\omega}{\omega_c})^2 \\
  -(\frac{\omega}{\omega_c})^3e^{-\frac{\omega}{\omega_c}}Ei(\frac{\omega}{\omega_c})\Big].
\end{array}$ \\
  \hline
\end{tabular}
\caption{\label{table1}Integral transforms between bath correlation function [$C(t)$], and spectral functions $\gamma(\omega)$ (spectral density) and $S(\omega)$ (principal density) at zero temperature.
 $\mathcal{P}$ indicates the principal value of the integral.
The second and third row display these functions in case of an Ohmic bath with exponential and Drude-Lorentz cutoff at frequency $\omega_c$.
Here, $\Theta(x)$ is the Heaviside step function, $Ei(x)$ is the exponential integral $\int_{-\infty}^x\frac{e^t}{t}dt$, $g$ is the dimensionless system-bath coupling constant, and $\gamma$ is the Euler–Mascheroni constant. The forth row exhibits the super-Ohmic spectral density.}
\end{table}

\section{\label{sec:RMA} Redfield equation}

The equation is a widely used fundamental master equation describing fluctuations and dissipation in quantal systems.~\cite{kubo}
It is derived from first principles, by applying the Born-Markov approximation and projection operator technique.
We begin here from Eq.~16 from a recent re-derivation,~\cite{mozgunov}
\begin{equation}
\frac{d\rho}{dt}=-i[H_0,\rho]-AA_f^\dagger \rho-\rho A_fA+A_f^\dagger\rho A +A\rho A_f,
\label{eq:redfield}
\end{equation}
where
\begin{equation}
\label{Eq:filter}
A_f=\int_0^\infty C(-\tau)A(-\tau)d\tau\equiv A\circ \Gamma^\star
\end{equation}
is termed the "filtered" operator, and adopt the notation from that reference throughout this paper.
The star symbol indicates complex-conjugate. The matrix elements of $\Gamma$ can be expressed as
\begin{equation}
\Gamma_{nm}=\int_0^\infty C(\tau)e^{i\omega_{nm}\tau}d\tau=\frac{1}{2}\gamma_{nm}+iS_{nm}.
\label{eq:filter1}
\end{equation}
In this work we also use functions
$\Gamma(x)=\frac{1}{2}\gamma(x)+iS(x)$, so that $\Gamma_{nm}=\Gamma(\omega_{nm})$, $\gamma_{nm}=\gamma(\omega_{nm})$, and $S_{nm}=S(\omega_{nm})$.
Here $\gamma(x)$ is the spectral density (SD), equal to the Fourier transform of the bath correlation function,  and  $S(x)$ is the principal density (PD), equal to the imaginary part of the half-range Fourier transform of the bath correlation function. The symbols $\Gamma$, $\gamma$, and $S$  mean matrices with respective matrix elements  $\Gamma_{nm}$,  $\gamma_{nm}$ and $S_{nm}$.
Table~\ref{table1} summarizes key formulae for the bath correlation functions and its integral transforms, and three examples of heat baths at zero temperature, that we apply in this paper.

Eq.~\ref{eq:redfield} can be recast in a different form,
\begin{equation}
\frac{d\rho}{dt}=-i[H_0,\rho]-\frac{1}{2}[AA_f^\dagger-A_fA,\rho]-
\frac{1}{2}\{AA_f^\dagger+A_fA,\rho\}+A_f^\dagger\rho A+A\rho A_f.
\end{equation}
With the renormalization of the system Hamiltonian,
\begin{equation}
H=H_0-\frac{i}{2}(AA_f^\dagger-A_fA)\equiv H_0+H_L,
\label{eq:shif}
\end{equation}
the Redfield equation becomes
\begin{equation}
\frac{d\rho}{dt}+i[H,\rho]=-\frac{1}{2}\{AA_f^\dagger+A_fA, \rho\}+A\rho A_f+A_f^\dagger\rho A,
\label{eq:redfieldsymmetrized}
\end{equation}
or in terms of matrix elements,
\begin{align}
\notag
\frac{d\rho_{nm}}{dt}+i[H,\rho]_{nm} =&
{\sum_{ij}} A_{in}^\star\rho_{ij}A_{jm}
\left[ \frac{\gamma_{in}+\gamma_{jm}}{2}-i(S_{jm}- S_{in} ) \right] \\
\notag
&-\frac{1}{2}{\sum_{i,j}}\rho_{ni}A_{mj}^\star A_{ij}\left[ \frac{\gamma_{ij}+\gamma_{mj}}{2}-i(S_{ij}- S_{mj}) \right]\\
\label{eq:RedDD}
&-\frac{1}{2}{\sum_{i,j}} A_{ji}^\star A_{ni}\rho_{jm}\left[ \frac{\gamma_{ni}+\gamma_{ji}}{2}-i( S_{ni}-S_{ji} ) \right].
\end{align}
The matrix elements of the renormalized Hamiltonian~\ref{eq:shif} are
\begin{equation}
\label{eq:LambDD}
H_{nm} = E_n\delta_{nm}+{\sum_{i}} {\mathcal H}(\omega_{ni},\omega_{mi})A_{ni}A_{im},
\end{equation}
where we introduce the kernel for the unitary component of the reduced system dynamics (from now on the unitary kernel):
\begin{equation}
\label{eq:LambDDkernel}
{\mathcal H}(\omega,\omega')=\frac{1}{2}\left [
S(\omega)+S(\omega')+i\frac{\gamma(\omega)-\gamma(\omega')}{2}
\right].
\end{equation}

The role of the renormalization is to rewrite the loss terms in the Redfield equation  so
that all three summands in Eq.~\ref{eq:RedDD} involve this one and only function of system frequencies,
\begin{equation}
G(\omega,\omega')=\frac{1}{2}\left[\gamma(\omega)+\gamma(\omega')\right]-i\left[S(\omega')- S(\omega)\right],
\label{eq:normR}
\end{equation}
that we refer to here as the dissipative kernel. That is to say, we reformulate the Redfield equation as
\begin{align}
\notag
\frac{d\rho_{nm}}{dt}+i[H,\rho]_{nm}=& {\sum_{ij}} A_{in}^\star\rho_{ij}A_{jm}
G_{in,jm} \\
\label{eq:GRED}
&-\frac{1}{2}{\sum_{i,j}}\rho_{ni}A_{ij}A_{mj}^\star G_{mj,ij}
-\frac{1}{2}{\sum_{i,j}} A_{ni} A_{ji}^\star\rho_{jm}G_{ji,ni},
\end{align}
with one simple kernel
\begin{equation}
G_{in,jm}=G(\omega_{in},\omega_{jm})=\Gamma_{in}+\Gamma_{jm}^\star.
\label{eq:Gtensor}
\end{equation}

The importance of this form of the Redfield equation is that we can approximate the kernel only. The approximation will then carry over to the state gains and losses unbiasedly, maintaining the balance between them.
Furthermore, if the kernel in Eq.~\ref{eq:Gtensor} were positive-semidefinite, it would lead to a GKSL-equation.
See for example Ref.~\cite{Schaller}.
So restoring positivity of the Redfield equation now amounts to approximating its kernel with a positive-semidefinite one.

As written above, the Redfield equation has no dependence on the initial time and
the history of the state is fully encoded in the present state.
In the derivation of the Redfield equation, before the last approximation is applied, there is a master equation with time-dependent coefficients (TDCs), that depend on the choice of the initial time.
The filtered operator in that case has time dependence
\begin{equation}
\label{Eq:filtertdc}
A_f(t)=\int_0^t C(-\tau)A(-\tau)d\tau\equiv A\circ \Gamma^\star(t),
\end{equation}
which leads to the time-dependent SDs and PDs vis-{\`a}-vis
\begin{equation}
\Gamma_t(\omega_{nm}) = \frac{1}{2}\gamma_t(\omega_{nm})+iS_t(\omega_{nm}),
\label{eq:GammaTDC}
\end{equation}
while the master equation is
\begin{equation}
\frac{d\rho}{dt}=-i[H_0,\rho]-AA_f^\dagger(t) \rho-\rho A_f(t)A+A_f^\dagger(t)\rho A +A\rho A_f(t).
\label{eq:redfield-tdep}
\end{equation}
This equation
depends on the initial time,
but, as $t$ becomes much longer than $\tau_c$, the coefficients $\Gamma_{nm}(t)$ approach the asymptotic value,
and the equation no longer depends on the history of the system.~\cite{BreuerHeinz-Peter1961-2007TToO} We also call this the $\int^t\to\int^\infty$ approximation.

In this paper we refer to equations~\ref{eq:redfield-tdep} and~\ref{eq:redfield} as the TDC-Redfield equation and the Redfield equation, respectively.
The Redfield equation does not resolve the system dynamics over a time scale
of the bath correlation time.~\cite{BreuerHeinz-Peter1961-2007TToO}

A time-dependent Redfield equation in form equivalent to~\ref{eq:redfieldsymmetrized} has been studied recently in context of
embedding of the quantum system into an expanded system that displays completely positive
quantum dynamics by Breuer.~\cite{Breuer2004} The density matrix of the subsystem, after the embedding, is presented by the off-diagonal block of the expanded state,
and such embedding does not resolve negativity of the reduced state. The renormalization of the Hamiltonian is mentioned, but not written
down explicitly.

The range of applicability of the Redfield equation is given by the condition of the validity of the Born-Markov approximation.
The failure of the Born-Markov approximation can be identified by comparing the second and fourth order terms in the Dyson expansion of the system state,~\cite{Hone,Albash_2012} which leads to the condition $\tau_r\sim\tau_c$. The Born-Markov approximation is justified
if $\tau_r\gg\tau_c$.

For any given initial state of the system, the error bound of the Redfield equation can be expressed in terms of the properties of the heat bath only,~\cite{mozgunov}
\begin{equation}
\label{eq:redfieldError}
1/2\lvert\lvert \rho_{RED}(t)-\rho_{exact}(t)\rvert\rvert_{1}\leq
O\left(\frac{\tau_b}{\tau_{sb}}e^{\frac{12t}{\tau_{sb}}}\right)
\text{ln}\left(\frac{\tau_{sb}}{\tau_b}\right).
\end{equation}
Here $\rho_{exact}(t)$ is the unapproximated quantum state, $\tau_b$ and $\tau_{sb}$ are the bath correlation time and the smallest system relaxation time, respectively, and $\vert\vert ... \vert\vert_1$ is the trace-distance. $\tau_{sb}$ is roughly $\tau_b/g$. Ignoring the logarithmic correction, the error bound is linear in system-bath coupling constant $g$.
Recent simulations on a two qubit-system find linear scaling of the actual error (as opposed to the error bound) with $g$.~\cite{Hartmann}

In the numerical simulations in this paper, the state relaxation time will be significantly longer than $\tau_{sb}$, as it is in many cases in quantum optics and atomic physics.~\cite{girvin,Daley} In that case the error bound~\ref{eq:redfieldError} is exponentially large at the system relaxation time. Only if the system relaxation time
is comparable to $\tau_{sb}$ the bound will be tight enough to present a good error estimate.
So here we adopt a system-dependent criterion to determine the range of applicability of the Redfield master equation. If, for example, we find that the Redfield states become too negative, we consider it as a sign that we are outside the range of applicability. A similar point of view has been expressed in Ref.~\cite{Hartmann}. In addition, if the Redfield equation has an instability, which it usually does at large system-bath coupling, then it will clearly not be applicable.

\subsection{~\label{sec:CRE} Coarse-grained Redfield Equation}

Here we coarse-grain Eq.~\ref{eq:GRED}, as follows.
First, we transform to the interaction picture [$\varrho_{nm}=\exp(i\omega_{nm}t)\rho_{nm}$], where the Redfield equation reads
\begin{align}
\notag
\frac{d\varrho_{nm}}{dt}=&-i[e^{iH_0t}H_Le^{-iH_0t},\varrho]_{nm}+
{\sum_{ij}} A_{in}^\star\varrho_{ij}A_{jm}e^{i(\omega_{jm}-\omega_{in})t}G(\omega_{in},\omega_{jm})\\
\label{eq:RedDDINT}
&-\frac{1}{2}{\sum_{i,j}} \left[\varrho_{ni}A_{ij}A_{mj}^\star e^{i\omega_{im}t} G(\omega_{mj},\omega_{ij})+A_{ni}A_{ji}^\star\varrho_{jm}e^{i\omega_{nj}t}G(\omega_{ji},\omega_{ni})\right].
\end{align}
Next, we change the time symbol from $t$ to $\tau$, and time average on a rectangular window. That is,
we apply the integral operator to the equation,
\begin{equation}
\label{eq:hatG}
\hat{G}~ =
\frac{1}{T_0}\int_{t-T_0/2}^{t+T_0/2}d\tau
\end{equation}
while keeping $\varrho$ at time $t$. We will refer to $T_0$ as the coarse-graining time.
This leads to
\begin{align}
\notag
\frac{d\varrho_{nm}}{dt}+i[\tilde{H}_L(t),\varrho]_{nm} =&
{\sum_{ij}} A_{in}^\star\varrho_{ij}A_{jm}e^{i(\omega_{jm}-\omega_{in})t}{\tilde G}(\omega_{in},\omega_{jm})\\
\label{eq:RedDDcg}
&-\frac{1}{2}{\sum_{i,j}} \left[\varrho_{ni}A_{ij}A_{mj}^\star e^{i\omega_{im}t} {\tilde G}(\omega_{mj},\omega_{ij})+A_{ni}A_{ji}^\star\varrho_{jm}e^{i\omega_{nj}t}{\tilde G}(\omega_{ji},\omega_{ni})\right],
\end{align}
where
\begin{equation}
\label{eq:CGG}
{\tilde G}(\omega,\omega')=
\bigg\{\frac{1}{2}\left[\gamma(\omega)+\gamma(\omega')\right]
+i\left[S(\omega)- S(\omega')\right]\bigg\}\text{sinc}\frac{(\omega-\omega')T_0}{2}.
\end{equation}
Here $\text{sinc}(x)=[\sin (x)]/x$, and
\begin{equation}
\tilde{H}_L(t)={\hat G}~ e^{iH_0\tau}H_Le^{-iH_0\tau}.
\end{equation}

In the third step, we transform out of the interaction picture, [$\rho_{nm}=\exp(-i\omega_{nm}t)\varrho_{nm}$],
and obtain the coarse-grained Redfield equation,
\begin{align}
\notag
\frac{d\rho_{nm}}{dt}+i[\tilde{H},\rho]_{nm} =&
{\sum_{ij}} A_{in}^\star\rho_{ij}A_{jm}{\tilde G}(\omega_{in},\omega_{jm})\\
\label{eq:RedCG}&-\frac{1}{2}{\sum_{i,j}}\left[\rho_{ni}A_{ij}A_{mj}^\star {\tilde G}(\omega_{mj},\omega_{ij})+A_{ni}A_{ji}^\star\rho_{jm}{\tilde G}(\omega_{ji},\omega_{ni})\right].
\end{align}

Similarly, the matrix elements of the coarse-grained renormalized Hamiltonian are
\begin{equation}
\label{eq:lambCG}
\tilde{H}_{nm}\equiv E_n\delta_{nm}+ \tilde{H}_{L,nm}=E_n\delta_{nm}+{\sum_{i}} {\tilde{\mathcal H}}(\omega_{ni},\omega_{mi})A_{ni}A_{im},
\end{equation}
where
\begin{equation}
{\tilde{\mathcal H}}(\omega,\omega')={\mathcal H}(\omega,\omega')\text{sinc}\frac{(\omega-\omega')T_0}{2}.
\label{eq:lambCGkernel}
\end{equation}

Coarse-graining increases the error of the approximate state. Error bound exists
for the state error added by coarse-graining, and increases linearly with the coarsegraining time.~\cite{mozgunov}

\subsection{\label{sec:RWA}Rotating wave and CGSE approximations}

The RWA follows from Eqs.~\ref{eq:CGG}, \ref{eq:RedCG} and~\ref{eq:lambCG}
in the limit $T_0\gg 1/\lvert\omega_{nm}\rvert$,  $\forall\omega_{nm}\neq 0$. In that limit the sinc-function becomes the Kronecker-delta thereby truncating the coarse-grained Redfield equation.
The truncation causes cancellations of the principle densities $S$ on the RHS of Eq.~\ref{eq:RedDD}, leading to
the historically important Davies-Lindblad master equation:
\begin{equation}
\label{eq:davies00}
\frac{d\rho_{nm}}{dt}
+i[\tilde{\tilde{H}},\rho]_{nm}=
\sum\limits_{i,j}\delta_{\omega_{in},\omega_{jm}}\gamma_{in}A_{in}^\star\rho_{ij}A_{jm}
-\frac{1}{2}\sum\limits_{ij}\gamma_{ij}\left(\rho_{ni}A_{ij}A_{jm}\delta_{E_i,E_m}
+A_{nj}A_{ji}\delta_{E_n,E_i}\rho_{im}\right)
\end{equation}
where $\tilde{\tilde{H}}$ is the system Hamiltonian renormalized due to the environmental coupling,
\begin{equation}
\tilde{\tilde{H}}_{nm}=E_n\delta_{n,m}+\sum\limits_i \tilde{\tilde{\mathcal{H}}}(\omega_{ni},\omega_{mi})A_{ni}A_{im},
\label{eq:lambDavies00}
\end{equation}
which is also known as the Lamb-shift. The unitary kernel in the RWA is
\begin{equation}
\label{eq:RWAkernel}
\tilde{\tilde{\mathcal{H}}}(\omega,\omega') = \frac{\delta_{\omega,\omega'}}{2}\left[S(\omega)+S(\omega')\right].
\end{equation}

In the interaction picture, the Davies-Lindblad master equation has time independent coefficients. The equation and the Lamb-shift simplify further, respectively as
\begin{equation}
\label{eq:davies}
\frac{d\rho_{nm}}{dt}=
-i\epsilon_{nm}\rho_{nm}+\sum\limits_{i,j}\delta_{\omega_{in},\omega_{jm}}\gamma_{in}A_{in}^\star\rho_{ij}A_{jm}-\frac{1}{2}\rho_{nm}\sum\limits_{i}\left(\lvert A_{ni}\rvert^2\gamma_{ni}
+\lvert A_{mi}\rvert^2\gamma_{mi}\right),
\end{equation}
and
\begin{equation}
\label{eq:lambDavies}
\epsilon_n=E_n+\sum_i\vert A_{n,i}\vert^2 S_{n,i},
\end{equation}
if the energy levels of the system are non-degenerate.

RWA is valid when the coarse-graining time is longer than the Heisenberg time $\tau_h=1/\delta E$, where $\delta E$ is the smallest level spacing in the system.
In that respect, the system relaxation time also needs to be of that order
or longer, so that the coarse-grained equation can be of value. Therefore, the main disadvantage to the RWA is in its highly limited range of applicability.
Note that at zero temperature, the dissipative part of the system dynamics
under the RWA only involves positive frequencies that correspond to the emission of quanta from the system into the bath (since the SD at negative frequency is zero).~\cite{girvin}

More recently, a coarse-grained stochastic equation was derived~\cite{Majenz} with better error
and range of applicability than that of Davies',
\begin{equation}
\label{eq:CGSEerror}
1/2\lvert\lvert\rho_{cgse}(t)-\rho_{exact}(t)\rvert\rvert_{1}\leq O\left(\sqrt{\frac{\tau_c}{\tau_{sb}}}\right),
\end{equation}
but not as low as the Redfield error bound given by Eq.~\ref{eq:redfieldError}.
\pagebreak

\section{\label{sec:game}Geometric Arithmetic Master Equation (GAME)}

Since coarse-graining increases the approximate state error, as discussed at the end of Sec.~\ref{sec:CRE}, here we will attempt to restore positivity by picking and choosing which part of the Redfield equation to coarse-grain and by how much.
First, we
rewrite the Redfield equation (Eq. ~\ref{eq:RedDD}) in terms of the geometric mean SD, instead of the arithmetic one. For example, we write
$(\gamma_{in}+\gamma_{jm})/2=\sqrt{\gamma_{in}\gamma_{jm}}+(\sqrt{\gamma_{in}}-\sqrt{\gamma_{jn}})^2/2$. The
Redfield equation~\ref{eq:RedDD} becomes
\begin{align}
\notag
\frac{d\rho_{nm}}{dt}=
&-i[H,\rho]_{nm}+\sum_{i,j}\left( \rho_{ij}A_{in}^\star A_{jm}\sqrt{\gamma_{in}\gamma_{jm}}
-\frac{1}{2}\rho_{ni}A_{ij}A_{mj}^\star\sqrt{\gamma_{ij}\gamma_{mj}}
-\frac{1}{2}A_{ji}^\star A_{ni}\rho_{jm}\sqrt{\gamma_{ji}\gamma_{ni}}\right)\\
\label{eq:SRedDD}
&+\sum_{ij}\left[ \rho_{ij}A_{in}^\star A_{jm} f\left(\omega_{in},\omega_{jm}\right)-
\frac{1}{2}\rho_{ni}A_{ij}A_{mj}^\star f(\omega_{ij},\omega_{mj})
-\frac{1}{2}A_{ji}^\star A_{ni}\rho_{jm} f(\omega_{ji},\omega_{ni})
\right],
\end{align}
where we introduce the "detuning" function,
\begin{equation}
\label{eq:errorR1}
f(\omega,\omega')= \frac{1}{2}\left[\sqrt{\gamma(\omega)}-\sqrt{\gamma(\omega')}\right]^2+i\left[S(\omega)-S(\omega')\right].
\end{equation}
Note that $f(\omega,\omega)=0$, which is why we call it the detuning function. We can write
$\omega=\omega_a+(\omega-\omega')/2$ and $\omega'=\omega_a-(\omega-\omega')/2$, where $\omega_a=(\omega+\omega')/2$ is the center frequency and $\omega-\omega'$ is the detuning. For small detuning, defined as $\vert\omega-\omega'\vert<\vert\omega_a\vert$, we apply the Tyler expansion and find
\begin{equation}
f(\omega,\omega')=i(\omega-\omega')\frac{\partial S(\omega_a)}{\partial\omega_a} +O\left[(\omega-\omega')^2\right].
\label{Eq:detuninTyler}
\end{equation}

\begin{wrapfigure}{l}{0.65\textwidth}
\centering
\includegraphics[width=0.65\textwidth]{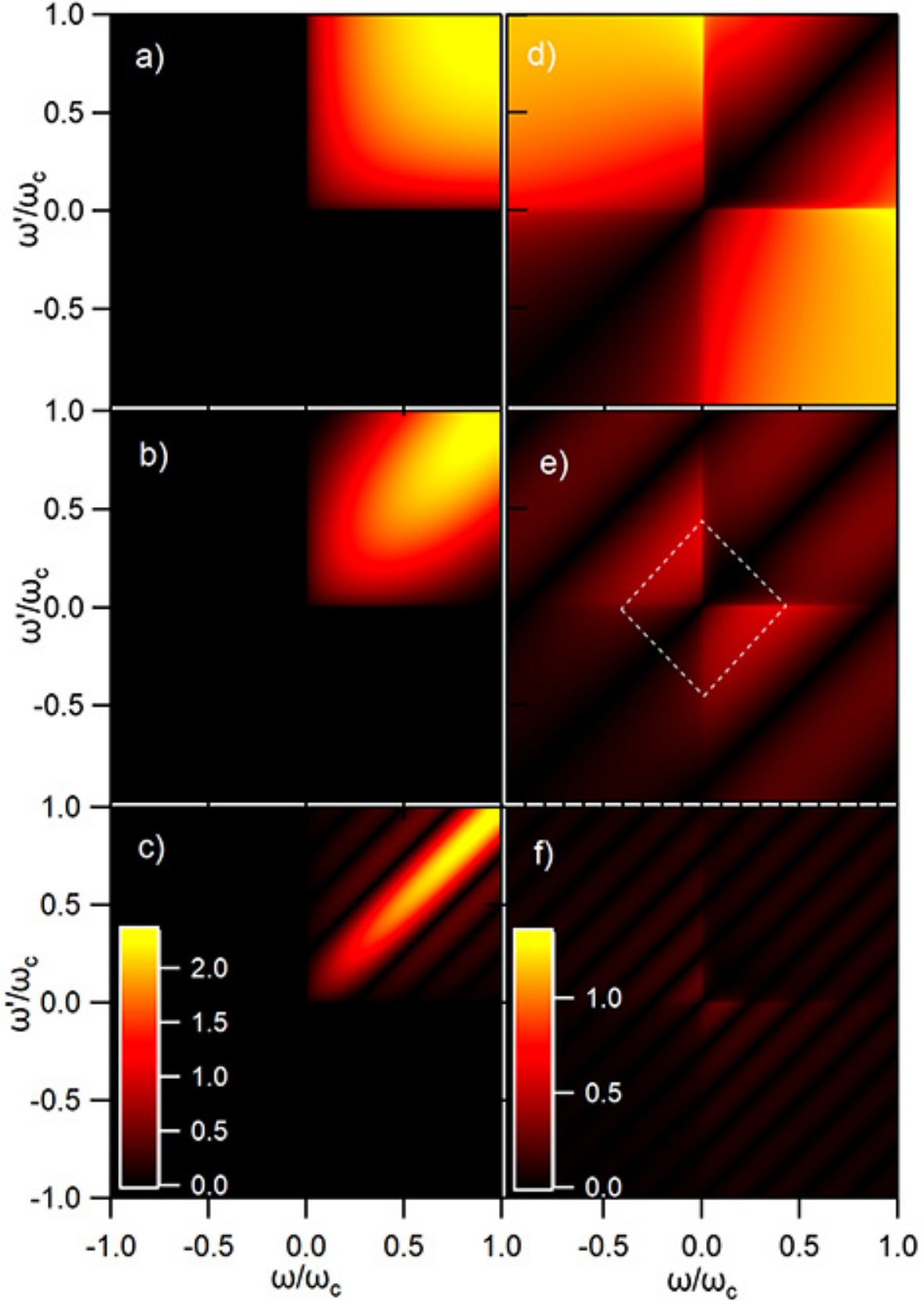}
\caption{a)\label{fig:ranges} The effect of coarse-graining. a)-c) Coarse-grained geometric mean SD, at the respective coarse-graining times corresponding to $\omega_cT_0=0$, $6.8$, and $25$. Yellow-Hot scale range is the same between the panels. d)-f) Coarse-grained detuning function, at $\omega_cT_0=0$, $6.8$, and $25$, respectively. Yellow-Hot scale range is the same between the panels, but reduced by factor of two relative to that in panels a)-c). $g=1$.}
\end{wrapfigure}

For Ohmic bath at zero temperature and exponential cutoff, the images in Fig.~\ref{fig:ranges}(a) and (d) display the geometric mean $g(\omega,\omega')=\sqrt{\gamma(\omega)\gamma(\omega')}$ and the detuning function magnitude versus system frequencies, respectively. The entities $f$ and $g$ are independent of the system operator $A$.
Without any coarse-graining, $f$ and $g$ are overall comparable in magnitude, which is unfortunate, because,
if the detuning function were negligibly small compared to the geometric mean, we could neglect the second line in Eq.~\ref{eq:SRedDD}, and end up with a GKSL master equation.

In spite of the significant value of the detuning function, we can make a compromise following the steps discussed next.
We recast the coarse-grained Eq.~\ref{eq:RedCG} as

\begin{align}
\notag
\frac{d\rho_{nm}}{dt}&=
-i[\tilde{H},\rho]_{nm}\\
\notag
&+\sum_{i,j}\left[ \rho_{ij}A_{in}^\star A_{jm}\tilde{g}(\omega_{in},\omega_{jm})
-\frac{1}{2}\rho_{ni}A_{ij}A_{mj}^\star\tilde{g}(\omega_{ij},\omega_{mj})
-\frac{1}{2}A_{ji}^\star A_{ni}\rho_{jm}\tilde{g}(\omega_{ji},\omega_{ni})\right]\\
\label{eq:SCGRedDD}
&+\sum_{ij}\left[ \rho_{ij}A_{in}^\star A_{jm} ~\tilde{f}\left(\omega_{in},\omega_{jm}\right)-
\frac{1}{2}\rho_{ni}A_{ij}A_{mj}^\star ~\tilde{f}(\omega_{ij},\omega_{mj})
-\frac{1}{2}A_{ji}^\star A_{ni}\rho_{jm} ~\tilde{f}(\omega_{ji},\omega_{ni})
\right],
\end{align}
where  $\tilde{g}$ and $\tilde{f}$ are the coarse-grained geometric mean and detuning function,
\begin{equation}
\tilde{g}(\omega,\omega')=\sqrt{\gamma(\omega)\gamma(\omega')}\,\text{sinc}\frac{(\omega-\omega')T_0}{2},
\label{eq:cggeometric}
\end{equation}
\begin{equation}
\tilde{f}(\omega,\omega')=\Big\{\frac{1}{2}\left[\sqrt{\gamma(\omega)}-
\sqrt{\gamma(\omega')}\right]^2+i\left[S(\omega)-S(\omega')\right]\Big\}\text{sinc}\frac{(\omega-\omega')T_0}{2}.
\label{eq:cgdetuning}
\end{equation}

The characteristic detuning frequency of these functions is governed by the sinc function. Irrespective of the center frequency, if $\vert\omega-\omega'\vert\gg 1/T_0$, then the sinc function will suppress
$\tilde{f}$ in Eq.~\ref{eq:cgdetuning} by  prefactor of $1/T_0$.

Lets us next consider the other frequency range, $\vert \omega-\omega'\vert < 1/T_0$.
If in this range $\vert\omega_a\vert\gg 1/T_0$, then the condition $\vert\omega-\omega'\vert\ll\vert\omega_a\vert$ will hold.
Eq.~\ref{Eq:detuninTyler} is then applicable and leads to
\begin{equation}
\begin{array}{cl}
&\tilde{f}(\omega,\omega')\approx i(\omega-\omega')\\
&\times\frac{\partial S(\omega_a)}{\partial\omega_a}\text{sinc}\frac{(\omega-\omega')T_0}{2}\\
&\sim \frac{2i}{T_0}\frac{\partial S(\omega_a)}{\partial\omega_a}\text{sinc}\frac{(\omega-\omega')T_0}{2}.
\end{array}
\end{equation}
So the corresponding terms in Eq.~\ref{eq:SCGRedDD} are also suppressed by the inverse coarse-grain time.
Vice-versa, if $\vert\omega_a\vert<1/T_0$, then Eq.~\ref{Eq:detuninTyler} may not hold, and the corresponding terms in Eq.~\ref{eq:SCGRedDD} are not suppressed.
Putting it all together, $\vert\tilde{f}(\omega,\omega')\vert$ is not suppressed in the frequency range
$\vert\omega\pm\omega'\vert<1/T_0$, which corresponds to a region bordered by a rotated square in the frequency space [Fig.~\ref{fig:ranges}(e)].
As the coarsegraining time increases, the square shrinks towards the origin and the entire function $\tilde{f}(\omega,\omega')$ is suppressed.

Let us visually inspect these coarse-graining effects at times $T_0=6.8/\omega_c$ and $25/\omega_c$.
The coarse-grained geometric mean SD, displayed in Figs.~\ref{fig:ranges}(b,c), is significantly suppressed in the frequency range $\lvert\omega-\omega'\rvert\ > 1/T_0$, but remains mostly unchanged at frequencies near the diagonal $\omega=\omega'$.  The coarse-grained detuning function is similarly suppressed in the frequency range $\lvert\omega-\omega'\rvert> 1/T_0$, as shown in Figs.~\ref{fig:ranges}(e,f). In contrast to the geometric mean, the coarse-grained detuning function is strongly suppressed in its entirety. The white-dashed rotated square in Fig.~\ref{fig:ranges}(e) is for illustrative purpose only, displaying the region inside which the detuning function is not suppressed by coarse-graining. The square shrinks as $T_0$ increases. An important point to make here is that the characteristic coarse-graining time that suppresses the detuning function relative to the geometric mean is independent of the strength of the coupling of the system to the environment and of the details of the system operator $A$.

\begin{figure}
  \centering
    \includegraphics[width=0.85\textwidth]{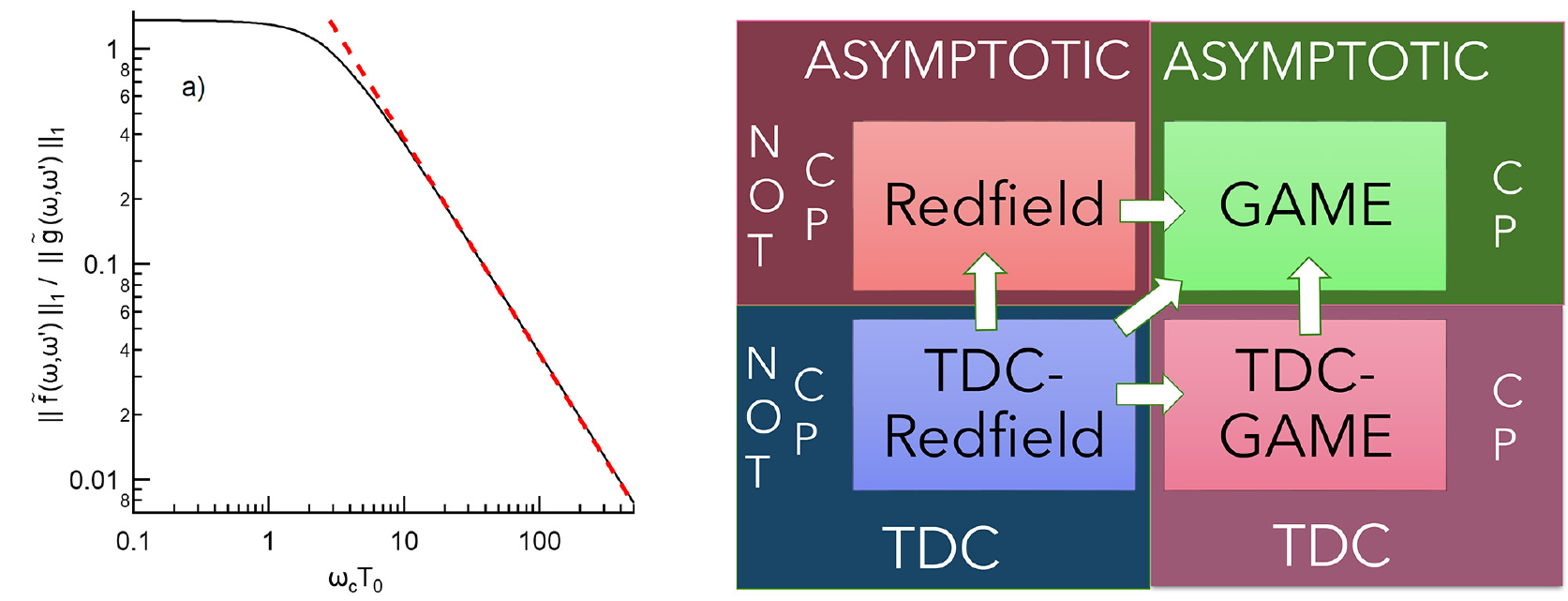}
  \caption{a) \label{fig:norma} Ratio of the tracenorms of two kernels versus coarse-graining time. Black-full curve is the ratio between the detuning-function and the geometric mean SD. Red-dashed line is the best inverse fit at large $T_0$. Diagram: Four master equations divided into quadrants based on complete positivity (CP) and time dependence of the coefficients.
GAME is initially obtained by approximating the Redfield equation. Its accuracy is assessed by comparison with the TDC-Redfield equation, eider directly or via an intermediate TDC-GAME equation.}
\end{figure}

For the Ohmic bath with exponential frequency cutoff, we quantify the effect of coarse-graining in terms of the trace-norm of functions $~\tilde{g}$ and $~\tilde{f}$. 
Fig.~\ref{fig:norma} displays the ratio of the norms of the coarse-grained detuning function and geometric mean.
While at small $T_0$, the norm ratio is constant, above a characteristic time it fits well to inverse time dependence as shown by the dashed line
in Fig.~\ref{fig:norma}.
There is a crossover time of order bath-correlation time $1/\omega_c$ at which the
time dependence changes from constant to inverse.

This analysis conveys a picture of dissipative quantum dynamics. The effects of the detuning function average out relative to the effects of the geometric mean density, on timescale governed by $\tau_c$, and possibly a system time-scale encoded in the operator $A$.
Namely, the second line of the master equation~\ref{eq:SCGRedDD} has a crossover coarse-graining time $T_C$ above which its contribution starts decreasing as inverse coarse-graining time. Then if we consider system dynamics on timescale longer than $T_c$, we can neglect the detuning function altogether,
but retain all other contributions in their primitive (e.g., not coarse-grained) form.

Next we introduce the generator matrix,
\begin{equation}
\label{eq:ansatz}
L=A\circ \sqrt{\gamma},
\end{equation}
(where $\sqrt{\gamma}$ is defined as the matrix with elements $\sqrt{\gamma_{ij}}$), which leaves us with a GKSL master equation that we call GAME,
\begin{equation}
\label{eq:incRWA1}
\frac{d\rho}{dt}=-i[H,\rho] -\frac{1}{2}\{LL^\dagger,\rho\}+L^\dagger\rho L,
\end{equation}
that is more general than the Davies-Lindblad equation.
Without the renormalized Hamiltonian $H$,
this equation would be identical to the phenomenological PERLind master equation.~\cite{perlind}
It is interesting that the coarse-graining time is absent in Eq.~\ref{eq:incRWA1}, just as it is absent in the Davies-Lindblad equation.

In this paper we study by how much the solutions of different master equations can diverge from each other for the same initial condition.
In the diagram in Fig.~\ref{fig:norma}, the Redfield and GAME are shown in the upper row. They both have asymptotic coefficients,
but differ in complete positivity. As described above we restore complete positivity of the Redfield equation
directly, by dropping the detuning function, which is the approximation step indicated by the arrow in the upper row of the diagram.

We also include the TDC-Redfield equation into the analysis, e.g.,
Eq.~\ref{eq:redfield-tdep},
because it is a higher level of approximation relative to the Redfield equation, in terms of accuracy.~\cite{mozgunov,Hartmann}
So we can directly compare GAME and TDC-Redfield equation, along the diagonal arrow in the diagram in Fig.~\ref{fig:norma}.

Alternatively, we first restore complete positivity of the TDC-Redfield equation by an arithmetic geometric mean SD approximation. This leads to an intermediate master equation with TDCs that we refer to as TDC-GAME.
Specifically, we introduce a time-dependent renormalization of the system Hamiltonian,
\begin{equation}
H(t)=H_0-\frac{i}{2}[AA_f^\dagger(t)-A_f(t)A]\equiv H_0+H_L(t).
\label{eq:shif-tdep}
\end{equation}
The matrix elements of $H(t)$ are expressed in analogy with the time-independent case,
\begin{equation}
\label{eq:LambDDt}
H_{nm}(t) = E_n\delta_{nm}+{\sum_{i}} {\mathcal H}_t(\omega_{ni},\omega_{mi})A_{ni}A_{im},
\end{equation}
where
\begin{equation}
\label{eq:LambDDtkernel}
{\mathcal H}_t(\omega,\omega')=\frac{1}{2}\left [
S_t(\omega)+S_t(\omega')+i\frac{\gamma_t(\omega)-\gamma_t(\omega')}{2}
\right].
\end{equation}

Now we rewrite Eq.~\ref{eq:redfield-tdep} analogous to how we rewrite the Redfield equation in Sec.~\ref{sec:RMA}.
Then we separate the geometric mean of the time-dependent SDs, and drop the time-dependent detuning function.
The resulting TDC-GAME equation in the Schr\"odinger picture is
\begin{equation}
\label{eq:incRWA1t}
\frac{d\rho}{dt}=-i[H(t),\rho] -\frac{1}{2}\{L(t)L(t)^\dagger,\rho\}+L(t)^\dagger\rho L(t),
\end{equation}
where in terms of matrix elements $\gamma_t(\omega_{nm})$ defined in Eq.~\ref{eq:GammaTDC}, we have
\begin{equation}
\label{eq:ansatzt}
L(t)=A\circ \sqrt{\gamma_t}.
\end{equation}
Notice that the matrix elements $\gamma_t (\omega_{nm})$ can be negative and therefore $\sqrt{\gamma_t}$ can be imaginary. Nevertheless Eq.~\ref{eq:incRWA1t}
retains the Lindblad form.  GAME is now derived from TDC-GAME,  by applying the $\int^t\to\int^\infty$ approximation, that is, by replacing the TDCs with their asymptotic values. We will study the relation between these four equations in Sec.~\ref{sec:TDC}.


\section{\label{sec:3levsys}3-level Jaynes-Cummings model}

Quantum correlations mediated by the heat bath can be exploited to generate
nonlocal unitary transformations including entanglement between qubits.~\cite{Schaller,Benatti_2009,Benatti,Majenz,Rivas,Hartmann} GAME can well describe such correlations. This is primarily due to the unitary component of the system evolution, governed by the renormalized Hamiltonian given by Eq.~\ref{eq:shif}. The correlations become particularly strong near the crossings of system frequencies, which turn to avoided crossings when the frequency difference becomes lower than $1/\tau_r$ as we study in this section.

We apply GAME on an exactly solvable Jaynes-Cummings type of model, complex enough to display these quantum correlations. Specifically, we utilize a V-type 3-level system studied in Ref.~\cite{Majenz}.

The unperturbed system eigenstates are $\vert 0\rangle$, $\vert 1\rangle$, and $\vert 2\rangle$,
with the respective sorted eigenenergies $E_0=0$, $E_1$, and $E_2$.
The total Hamiltonian of the system and the bath of noninteracting linear harmonic oscillators, is
\begin{equation}
H_{tot}=\sum_{i=0}^2E_i\lvert i\rangle\langle i\rvert+\sum_k\omega_k b^\dagger_k b_k+H_{sb},
\end{equation}
where $H_{sb}$ is the system-bath coupling Hamiltonian
\begin{equation}
\label{eq:3levs1}
H_{sb}=C\otimes B +C^\dagger \otimes B^\dagger.
\end{equation}
Here
\begin{equation}
\label{eq:3levs2}
C=\vert 1\rangle\langle 0\vert+\vert 2\rangle\langle 0\vert=\left(
\begin{array}{ccc}
0 & 0 & 0 \\
1 & 0 & 0   \\
1 & 0 & 0 \\
\end{array}
\right),
\end{equation}
\begin{equation}
B=\sum_k g_kb_k,
\label{eq:B}
\end{equation}
$b_k$ are boson annihilation operators, and $\omega_k\geq 0$.
The appropriate bath-correlation function
for this system is $C(t)=Tr[\rho_b B(t)B^\dagger(0)]=\sum_k \lvert g_k\rvert^2 e^{-i\omega_kt}$.~\cite{Majenz}

\begin{wrapfigure}{L}{0.56\textwidth}
\centering
\includegraphics[width=0.56\textwidth]{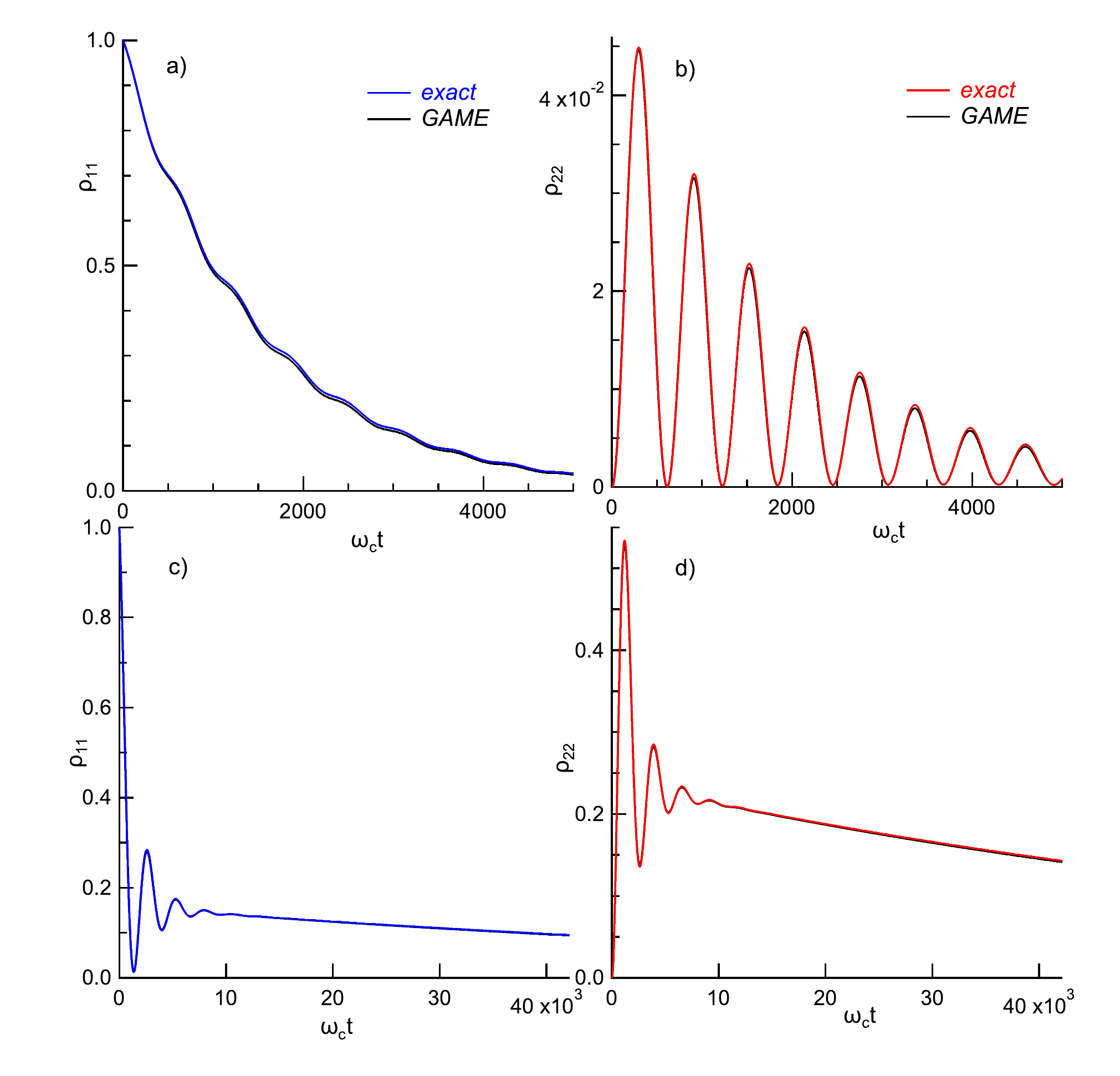}
\caption{Time dependence of the populations, for initial condition
$\rho(0)=\lvert 1\rangle\langle 1\rvert$. Blue and red lines are from the
exact solution. Black lines are from the approximate solution of Eq.~\ref{eq:incRWA1}.
a) and b): Case A: $E_1=0.095$, and $E_2=0.105$. c) and d): Case B: $E_1=0.09975$, and $E_2=0.10025$. $g=0.001$.
\label{fig:exactAB}}
\end{wrapfigure}

The Redfield master equation is
\begin{equation}
\frac{d\rho}{dt}=-i[H_0,\rho]-CC_f^\dagger\rho-\rho C_fC^\dagger+C_f^\dagger\rho C+C^\dagger\rho C_f,
\label{eq:redfield2}
\end{equation}
where
$C_f=C\circ (\gamma/2-iS)$.
Note that the Hermitian conjugate of $C$ is now explicit, in contrast to that in Eq.~\ref{eq:redfield} where it would be redundant.
Similarly, the renormalized Hamiltonian features the Hermitian conjugate of $C$
\begin{equation}
H=H_0-\frac{i}{2}(CC_f^\dagger-C_f C^\dagger),
\label{shift3lev}
\end{equation}
or explicitly
\begin{equation}
H=\left(
\begin{array}{ccc}
0 & 0 & 0 \\
0 & E_1+S(E_1) & {\overline S}- \frac{i}{4}\Delta \gamma   \\
0 &{\overline S} +\frac{i}{4}\Delta \gamma&  E_2+S(E_2) \\
\end{array}
\right),\label{eq:matrix}
\end{equation}
where ${\overline S}=\frac{1}{2}[S(E_1)+S(E_2)]$ and $\Delta\gamma = \gamma(E_2)-\gamma(E_1)$.
Lastly, the generator in equation~\ref{eq:incRWA1} has matrix elements
$L_{nm}=C_{nm}\sqrt{\gamma_{nm}}$.

For the initial condition, we suppose that the system is in state $\vert 1\rangle$.
Repeating from Ref.~\cite{Majenz}, the time dependent state of the total system in the interaction picture is
\begin{align}
\notag
\lvert\psi(t)\rangle&=\left[c_1(t)\lvert 1\rangle+c_2(t)\lvert 2\rangle\right]\otimes\lvert 0\rangle_b\\
\label{{eq:psitot3}}
&+\lvert 0\rangle\otimes \sum_k d_k(t)b^\dagger_k\lvert 0\rangle_b,
\end{align}
where $\lvert 0\rangle_b$ is the bath vacuum.
The exact solution follows by numerically solving
a system of two Volterra integro-differential equations,
\begin{equation}
\label{eq:c2dot}
\dot{c}_1=-f_1\star c_1-e^{i\omega_{12}t}f_2\star c_2,
\end{equation}
\begin{equation}
\label{eq:c3dot}
\dot{c}_2=-f_2\star c_2-e^{-i\omega_{12}t}f_1\star c_1,
\end{equation}
where $f_1(t)=\exp(i\omega_{10}t)C(t)$, $f_2(t)=\exp(i\omega_{20}t)C(t)$,
$f\star g=\int_{0}^{t}f(\tau)g(t-\tau)d\tau$.

For the system parameters we use $g=0.001$, and $E_1=0.095\omega_c$, and $E_{2}=0.105\omega_c$ in case A,
and $E_1=0.09975\omega_c$ and $E_{2}=0.10025\omega_c$ in case B. These are chosen to be the
same parameters as in Ref.~\cite{Majenz}, so that we can directly compare the accuracy of the master equations.
Figure~\ref{fig:exactAB}
displays numerically obtained exact populations of levels 1 and 2 versus time, in parameter cases A and B, respectively.
The blue and red curves reproduce previous work.~\cite{Majenz}

\subsection{Accuracy of the approximate solution}

Next we investigate the error of the approximate state. We solve Eq.~\ref{eq:incRWA1}  as explained in appendix~\ref{sec:App1}.
The approximate populations of levels 1 and 2 versus time are displayed by the black lines in Fig.~\ref{fig:exactAB}.
The difference between the approximate and the exact solution is barely discernible. Not shown is the coherence
$\rho_{12}(t)$, which also agrees well with the exact solution.

\begin{figure}
\includegraphics[width=1\textwidth]{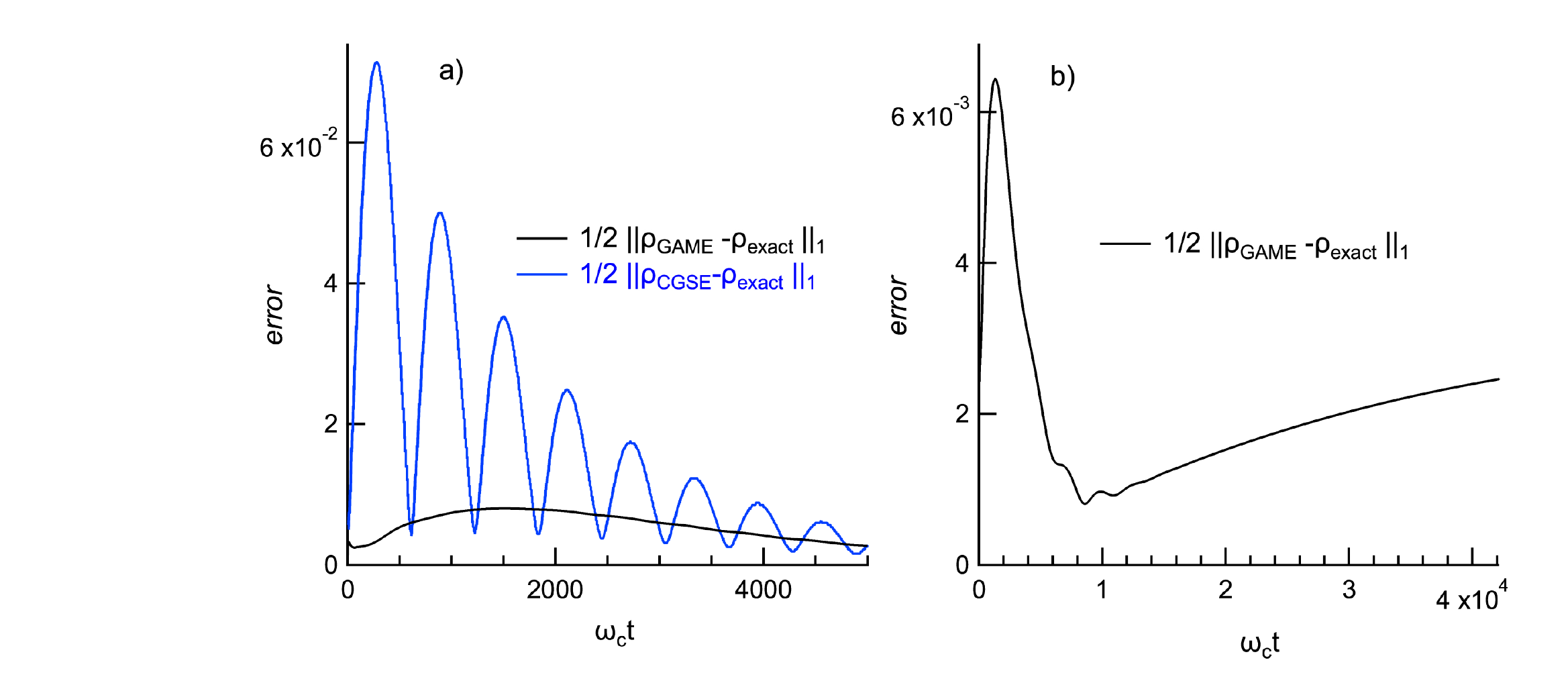}
\caption{\label{fig:errors}Trace-distance between the exact quantum state and the approximate states obtained
by solving the master equations. Black lines: Error of the solutions of the Lindblad equation (Eq.~\ref{eq:incRWA1}).
Blue line: Error of the optimal CGSE obtained by scanning and digitizing data in Ref.~\cite{Majenz}.
$g=0.001$. a) $E_1=0.095\omega_c$, $E_2=0.105\omega_c$. b) $E_1=0.09975\omega_c$, $E_2=0.10025\omega_c$.}
\end{figure}

The trace-distance between the approximate and exact solutions is displayed by the black lines in Fig.~\ref{fig:errors}.
Also shown by the blue line in Fig.~\ref{fig:errors}(a), is the error of the optimal CGSE, obtained by digitizing the blue line of data in Fig.~3 of
Ref.~\cite{Majenz}.  GAME is almost an order of magnitude more accurate than the CGSE, and almost completely captures the population oscillations as seen by the diffuse error with respect to time.

We also determine the trace-distance between the solutions of the Redfield and GAME equations,
and find that the distance is ten times lower than the black lines in Fig.~\ref{fig:errors}.
Additionally, we solve Eq.~\ref{eq:incRWA1} without renormalizing the system Hamiltonian
(Eq.~\ref{shift3lev}), which is equivalent to PERLind approach.~\cite{perlind} In that case the state error is much larger than the black lines in Fig.~\ref{fig:errors}. The error increases by a factor of several hundred and demonstrates the necessity to renormalize the Hamiltonian to properly account for quantum correlation.
The dissipator can induce correlations as well,~\cite{perlind} but it is insufficient for high accuracy. We will return to the subject of accuracy of PERLind in Sec.~\ref{sec:PERLind}.

Even though the interaction Hamiltonian (Eq.~\ref{eq:3levs1}) is of a RWA form, the Redfield equation in the interaction picture has oscillations and therefore it is not in the RWA. Although the large oscillation frequencies $\omega_{10}$ and $\omega_{20}$ are absent, the oscillations at detuning frequency $\omega_{21}$ remain.

The accuracy of the RWA in this system was found to be poor for the system parameters we apply.~\cite{Majenz} The error is exacerbated in the regime where the relaxation rate is larger than detuning $\omega_{21}$. This can be traced to a known property of the RWA~\cite{Benatti_2009,Benatti1,Hartmann} that there is an artificial discontinuity of the equation at zero detuning (level degeneracy). As an example, Eqs.~\ref{eq:lambDavies00} and~\ref{eq:lambDavies} do not merge at the crossings of levels $n$ and $m$.
In the three level system in particular, at finite detuning, the RWA of the Hamiltonian~\ref{eq:matrix} will be its diagonal part. At zero detuning, however, the RWA groups the frequency duplicates under same generator, leading to the
renormalized Hamiltonian~\ref{eq:matrix}.

The artifact has been remedied in case of two qubit systems coupled to a heat bath, by applying a coarse-grained approach,~\cite{Benatti_2009,Benatti1} which leads to a continuous description of entanglement creation as
function of the frequencies of the qubits. The entanglement between two-level atoms can also be created in the dissipative process of spontaneous emission.~\cite{Tana}
By similar mechanism, GAME replaces the discontinuity of the RWA with a continuous avoided crossing, as will be shown next, thereby creating superpositions between levels 1 and 2 at finite detuning.

\subsection{Results and discussion}

Interestingly, Figs.~\ref{fig:exactAB}~c) and d) show a few initial oscillations that decay rapidly followed by
a much slower relaxation process without any oscillations.
In case A, as can be seen in Figs.~\ref{fig:exactAB}~a) and b), the oscillations and populations decay on a similar scale.
At first sight, this property is puzzling since the Fermi-golden rule rates are approximately the same in cases A and B.

Thinking further, however, the disparity between the relaxation rates is not utterly surprising.
Namely, if levels $E_1$ and $E_2$ were degenerate, one could find a basis such that one state has a zero matrix element in the coupling Hamiltonian.
For example, if in one basis the coupling Hamiltonian is represented by the matrix in Eq.~\ref{eq:3levs2}, we can change the basis to
$\left[\lvert 0 \rangle,\,(\lvert 1\rangle+ \lvert 2\rangle )/\sqrt{2},\,(\lvert 1\rangle- \lvert 2\rangle)/\sqrt{2}\right]$.
$(\lvert 1\rangle- \lvert 2\rangle)/\sqrt{2}$ is the dark state.~\cite{Majenz}
If the levels are not exactly, but nearly degenerate, the bath will renormalize the states so that one state approaches the dark state
resulting in the suppressed relaxation rate.

Here we present a simple procedure that determines the system oscillation frequencies
and relaxation rates based entirely  on the properties of the renormalized Hamiltonian given by Eq.~\ref{eq:matrix}.
The renormalized energies are defined as the nonzero eigenvalues of the matrix in Eq.~\ref{eq:matrix},
\begin{equation}
E_{1,2}'=\overline{E}+\overline{S}\mp\sqrt{\overline{S}^2+\left(\frac{\Delta E+\Delta S}{2}\right)^2+\left(\frac{\Delta\gamma}{4}\right)^2},
\label{eq:quasienergies}
\end{equation}
where $\overline{E}=(E_1+E_2)/2$, $\Delta S=S(E_2)-S(E_1)$, and the other terms are identical to those in Eq.~\ref{eq:matrix}.
\begin{figure}
\includegraphics[width=1\textwidth]{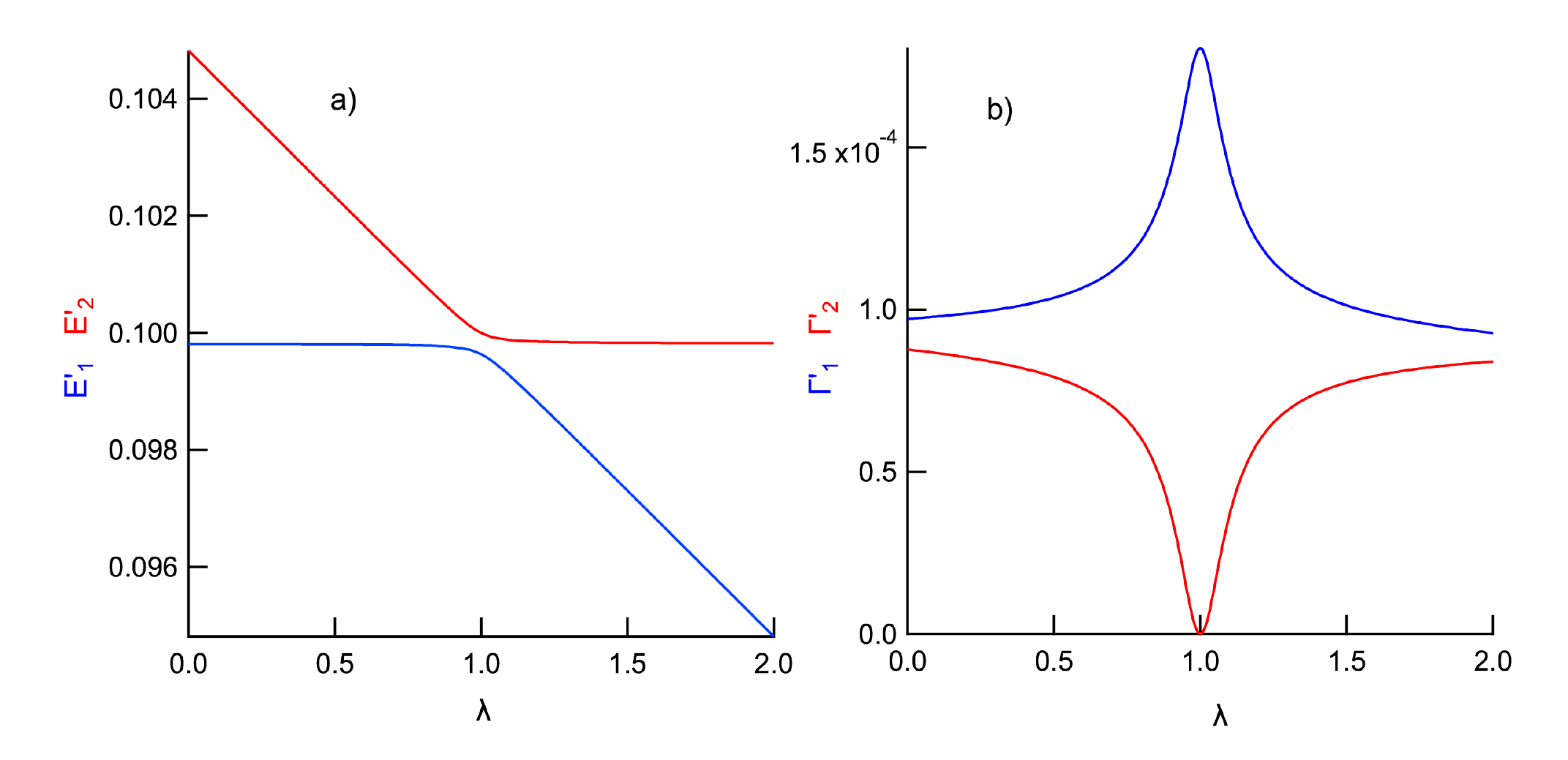}
\caption{Renormalized eigenenergies and relaxation rates in units of $\omega_c$, versus parameter $\lambda$ of the unperturbed Hamiltonian.
The higher level (red) has suppressed effective width and becomes a dark state with zero
relaxation rate at $\lambda=1$. The energy gap imposed by the heat bath is $2S(E_1)$, where $S$ is the PD.
At $\lambda=0$,  $E_1=0.1\omega_c$, $E_2=0.105\omega_c$.
\,$g=0.001$. \label{fig:cross}}
\end{figure}

\begin{figure}
\includegraphics[width=1\textwidth]{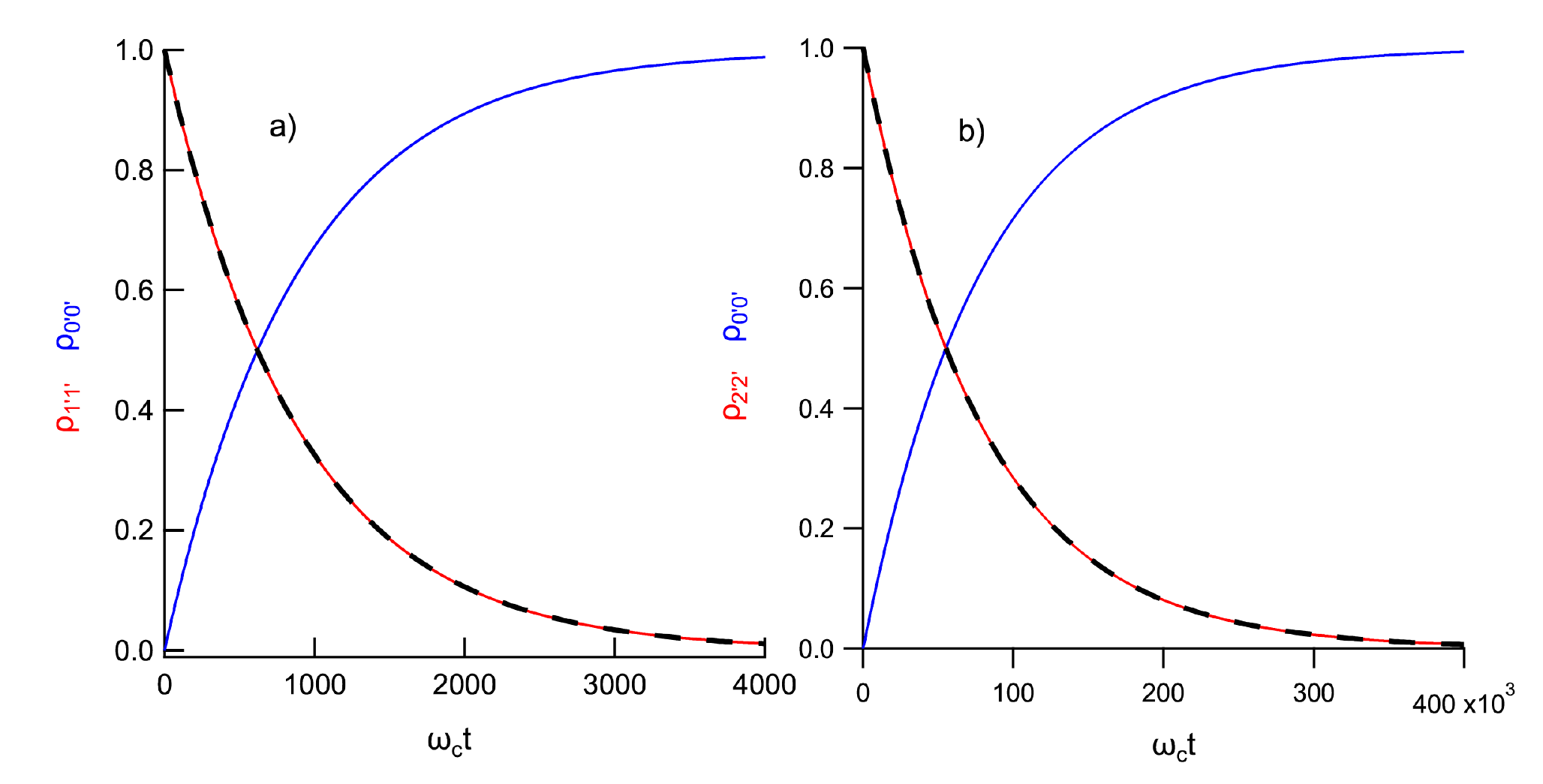}
\caption{Populations of the renormalized states versus time, in case B.
$E_1=0.09975\omega_c$, $E_2=0.10025\omega_c$, and $g=0.001$. The initial conditions are $\rho(0)=\lvert 1'\rangle\langle 1'\rvert$ and
$\rho(0)=\lvert 2'\rangle\langle 2'\rvert$ in a) and b), respectively.
Dashed black line is the best fit to $\exp(-\Gamma't)$.  $\Gamma '$ is the effective (renormalized) width. \label{fig:expo}}
\end{figure}

Let us introduce a parameter dependence of the system Hamiltonian,
\begin{equation}
H_0=E_{1}\lvert 1\rangle\langle 1\rvert +\left[E_2-\lambda(E_2-E_1)\right]\lvert 2\rangle\langle 2\rvert.
\end{equation}
As a function of $\lambda$, the energy levels of  $H_0$  cross at $\lambda = 1$, while the
renormalized eigenenergies (Eq.~\ref{eq:quasienergies}) have avoided crossing as shown in Fig.~\ref{fig:cross}(a) with the energy gap $2S(E_1)$.
The avoided crossing regime is found in the region of energy where the widths $\gamma(E_{1,2})$ are comparable to the spacing $E_2-E_1$.
In the regime of strong anticrossing, e.g., $\lvert E_2-E_1\rvert\ll \gamma(E_{1,2})$,
Eq.~\ref{eq:quasienergies} reduces to
\begin{equation}
 E'_1=\overline{E}-2\overline{S}-\kappa\frac{(\Delta E)^2}{\overline{S}},\,
 E'_2=\overline{E}+\kappa\frac{(\Delta E)^2}{\overline{S}},
\label{eq:gap1}
\end{equation}
where
\[
\kappa=\left[\frac{1}{16}\left(\frac{\partial\gamma}{\partial E}\right)^2+\frac{1}{4}\left(1+\frac{\partial S}{\partial E}\right)^2\right]_{\Delta E=0}.
\]
At the center of the crossing, the renormalized states are
\begin{equation}
\vert 1'\rangle  =\frac{\vert 1\rangle+\vert 2\rangle}{\sqrt{2}},
\end{equation}
\begin{equation}
\vert 2'\rangle  =\frac{\vert 1\rangle-\vert 2\rangle}{\sqrt{2}}.
\end{equation}
The last is the dark state.

Here, the avoided crossing is caused by the coupling between the system and the environment of linear harmonic oscillators.
We find that not only the states, but also the relaxation rates are strongly affected near such avoided crossings. To obtain the relaxation rates, we
consider the initial states of the system
$\rho (0)=\vert 1'\rangle\langle 1'\vert$ and $\vert 2'\rangle\langle 2'\vert$ and determine the populations
of the renormalized states versus time again by solving Eq.~\ref{eq:incRWA1}. The results are shown
in Fig.~\ref{fig:expo} for case B. In contrast to the time dependence of the unperturbed states shown
in Fig.~\ref{fig:exactAB}, there are no oscillations in Fig.~\ref{fig:expo}. The goodness of fit to $e^{-\Gamma' t}$ in Fig.~\ref{fig:expo}
indicates that the populations of
the renormalized states decay with single effective or renormalized widths.

The width $\Gamma'$ versus $\lambda$ is displayed in
Fig.~\ref{fig:cross}(b). The higher energy state
has a strongly suppressed width and becomes dark at $\lambda =1$, while the lower one has
enhanced width and increases in brightness by factor of two at $\lambda=1$.

Table~\ref{table2} displays the effective level widths of the reduced 3-level system, in the regime near the avoided crossings. The accuracy of the prefactor in the relaxation rate of the dark state is approximately 2-3\%.
Note the "sum rule" where the sum of the widths $\Gamma_2'+\Gamma_3'$ of the renormalized states is independent of the coupling strength.

\begin{table}

\centering

\begin{tabular}{|c|c|c|}
  \hline
  & level 1 & level 2 \\
  \hline
bare widths & $\Gamma_1=\gamma(E_1)$ & $\Gamma_2=\gamma(E_2)$ \\
  \hline
renormalized energies
   & $E'_1=\overline{E}-2\overline{S}-\kappa\frac{(\Delta E)^2}{\overline{S}}$ & $E'_2=\overline{E}+\kappa\frac{(\Delta E)^2}{\overline{S}}$ \\
   \hline
  renormalized widths & $\Gamma_{1}'=\Gamma_1+\Gamma_2-\Gamma_{2}'$ & $\Gamma_{2}'=\frac{\Gamma_1+\Gamma_2}{4}\left [ \frac{\Delta E}{2\overline{ S}(E)}\right]^2$ \\
  \hline
\end{tabular}
\caption{\label{table2}Bare and renormalized parameters in the strong coupling case. }
\end{table}

\pagebreak

\section{\label{sec:spinchain}Heisenberg Ferromagnetic Spin-Chain}

As a testbed for various master equations, here we consider a spin-chain of up to $n=25$ spin-1/2 particles. Each particle
couples to its nearest neighbor via the ferromagnetic exchange interaction
and to each other through the $r^{-3}$ magnetic dipole-dipole interaction.
This system has a rapidly increasing density of states with energy  and is rich in many-body phenomena, including, the energy gap (ferromagnetic resonance), collective excitations (standing spin-waves), and topological excitations (domain walls).
There is a corresponding plethora of relaxation processes, including Gilbert damping, emission of spin-waves, and domain wall motion, spanning wide range of relaxation times, that can be exponential in system size.
The focus here will be on relaxation of the density matrix in response to a macroscopic displacement of the magnetization, which is initially set perpendicular to the chain (easy) axis.

\subsection{Magnetic Hamiltonian}

The system is modeled by the Heisenberg model with dipole-dipole coupling,
\begin{equation}
\label{eq:heisenberg}
H_0=-J\sum_{i=1}^{n-1}\vec{S}_i\vec{S}_{i+1}-\epsilon_d\sum_{i=1}^{n-1}\sum_{j=i+1}^{n}\frac{3S_{z,i}S_{z,j}
-\vec{S}_i\vec{S}_{j}}{(j-i)^3}.
\end{equation}
The Hamiltonian commutes with $S_z=\sum_iS_{iz}$, but does not commute with  $\vec{S}^2$, where $\vec{S}=\sum_i\vec{S}_i$.
Here $n=25$ and the eigenvalues display Kramers degeneracy. In addition, the Hamiltonian is invariant with inversion of spins
about $i=(n+1)/2$. We use
the Kramers, $S_z$, and the inversion symmetry to reduce the Hamiltonian, and find the eigenvalues and eigenstates using the Lanzos algorithm on sparse matrices. Most simulations were done on a 16-core Intel Xeon processor and
192GB RAM, while the heaviest simulations in Sec.~\ref{sec:benchmark} were done on a 32-core AMD EPYC and 384GB RAM. The eigenstates have quantum numbers $S_z=-S,-S+1,...,S$ and are even or odd functions of spin. The states with $S_z<0$ are obtained by time-reversal operation on the states with $S_z>0$.

\begin{figure}
\includegraphics[width=1\textwidth]{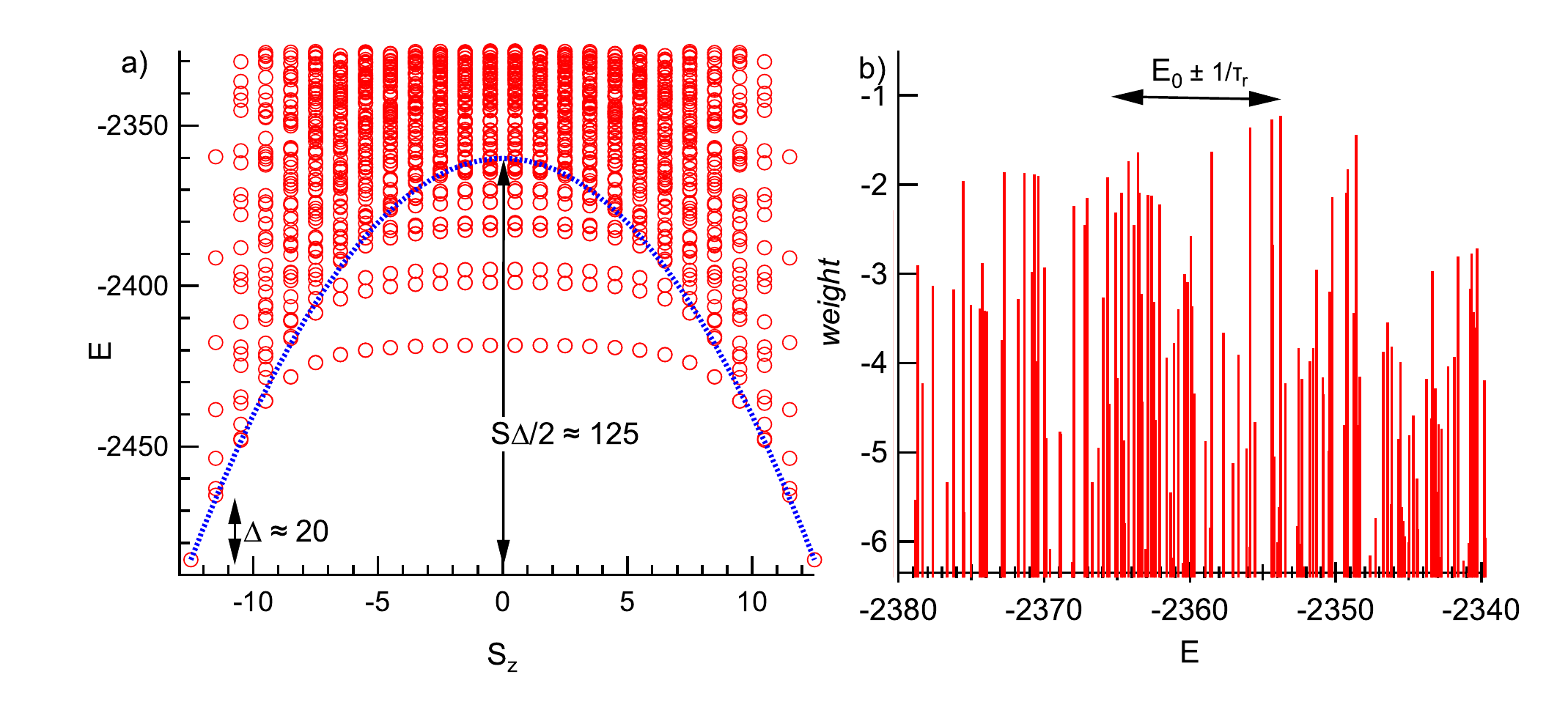}
\caption{a) Low lying energy levels in a 25 site spin-1/2 chain, in the Heisenberg model with short range ferromagnetic exchange and $1/r^3$
dipole-dipole interactions. $J=-400$ and $\epsilon_d=6$, respectively.
Blue parabola: Energy barrier between the ground states in the macrospin approximation.
b) Decomposition of the uniformly magnetized state ($\vert\Psi_x\rangle$) with magnetization oriented perpendicular to the easy axis of the spin-chain, in the energy eigenbasis. The weight on the y-axis is
defined as $\log \vert\langle E\vert\Psi_x\rangle\vert^2$. The black double-arrow line indicates the energy range centered at the top of the N{\' e}el barrier, at system-bath coupling $g_{tot}=1$. $n=25$ and $N=1728$.
\label{fig:spectrum}}
\end{figure}

For parameters we choose $J=400$ and $\epsilon_d=6$. With no anisotropy, (e.g., $\epsilon_d=0$), the spacing between the
lowest lying spin-multiplets
is  $\delta=2.5$. Each multiplet represents a different standing spin wave, while the spin-wave number varies within the corresponding multiplet.
In the case $\epsilon_d=6$, the ground state is that of a uniaxial ferromagnet, $\lvert S,S_z=\pm S\rangle$, (chain axis is along z-direction),
the ground state energy is $E_g=-2485.3$,
and there is an anisotropy gap  $\Delta=20.1$ in the excitation spectrum, as can be seen in Fig.~\ref{fig:spectrum}.
The gap is the ferromagnetic resonance. We work in the regime where $\Delta > \delta$, which is the regime common in devices.

Fig.~\ref{fig:spectrum} displays the lowest 1728 energy levels versus $S_z$. There are approximately $3\times 10^6$ Bohr frequencies spanning the range $-158.5<E_{ij}<158.5$.
In the macrospin approximation,
also known as the Stoner-Wohlfarth model,~\cite{Stoner} $H_{0,sf}=-KS_z^2/S$. The uniaxial anisotropy $K$ can be estimated from
the energy gap as $K=\Delta/2=10$. The blue parabola represents the energy of the magnet vs. $S_z$  in the
macrospin approximation. In that approximation, the N{\' e}el energy barrier for this uniaxial 1D magnet is $S\Delta/2=125$.

We are interested in the low energy collective magnetic dynamics. It suffices to find the eigenvalues and eigenstates,
spanning the energy range between the ground state to not far over the N{\' e}el  barrier, as we will see shortly.
In particular, let us consider the initial state with the magnetization fully oriented along the x-direction. To obtain this initial state, we truncate $S_x$ to the vector space on the lowest 1728 eigenstates and find
the eigenstate of the truncated $S_x$ with the maximum eigenvalue.
In that state, we obtain $S_x\approx 12.41$, which is lower than $12.5$ because of the truncation error.
Since this value is still very close to $12.5$, it shows that the vector space is sufficiently large to describe
the uniformly magnetized state, and the states accessible by relaxation from that state.
The expectation value of the energy in the perpendicularly magnetized state is approximately the
same as the energy at the top of the blue parabola in Fig.~\ref{fig:spectrum}(a),
while the energy uncertainty is $\text{rms}(H_0\dot{})=11.6$.

Note that we have applied a very weak magnetic field along the z-direction to break the $S_z$ degeneracy. The difference in energy between $\vert S,-S\rangle$ and $\vert S,S\rangle$ is $0.000125\Delta$.

The initial uniformly magnetized state is decomposed into the energy eigenstates as shown in Fig.~\ref{fig:spectrum}(b).
It shows a wide spread of the perpendicularly magnetized state over many eigenstates, with the spacing between different components much lower than the gap. Often these spacings will be smaller than the coupling to the heat bath,
which can introduce avoided crossings as we discussed in the example of the 3-level system. These properties make our system well suited for testing
completely positive master equations in many-body regime.

The energy levels under the parabola in Fig.~\ref{fig:spectrum}(a) are noteworthy. They are grouped into branches, which have weak dependence of energy with
$S_z$. On these branches, energies have two-fold near degeneracy [not visible in Fig.~\ref{fig:spectrum}(a)] between even and odd spin states. At
small $\lvert S_z\rvert$, these states are similar to domain walls located at distance $S_z$ from the middle of the chain.
The energy of the domain wall doesn't change much if it moves near the middle of the chain, explaining the flatness of the branches.
At this time, it would be a distraction to delve into these states but they may be interesting for future reasearch.

\subsection{Coupling to the heat bath}

Each spin is coupled to a noisy magnetic field, analogous to the environment in the celebrated Brown model.~\cite{brown}
In our case, there are 75 independent baths of linear harmonic oscillators, 3 for each spin, one for each direction of the field.
The coupling Hamiltonian is
\begin{equation}
H_{sb}=\sum_{i=1}^n\sum_{\alpha=x,y,z} S_{i\alpha}B_{i\alpha}.
\end{equation}
Here $B_{i,\alpha}=\sum_{k} g_k(b_{k,i\alpha}^\dagger+b_{k,i\alpha})$ are Hermitian bath operators, and we assume slow bath at zero temperature with Ohmic SD and cutoff frequency $\omega_c=6\Delta$. All baths have the same spectral functions. 

The Redfield equation for the spin-chain is
\begin{equation}
\frac{d\rho}{dt}=-i[H_0,\rho]+\sum_{i,\alpha}\left(-S_{i\alpha}S_{i\alpha,f}^\dagger \rho-\rho S_{i\alpha,f}S_{i\alpha}+S_{i\alpha,f}^\dagger\rho S_{i\alpha} +S_{i\alpha}\rho S_{i\alpha,f}\right),
\label{eq:redfieldChain}
\end{equation}
where $S_{i\alpha,f}=S_{i\alpha}\circ \Gamma$, and $\Gamma$ is given by Eq.~\ref{eq:filter1}.

The corresponding equation in GAME is
\begin{equation}
\frac{d\rho}{dt}=-i[H',\rho]+\sum_{i,\alpha}\left(L_{i\alpha}^\dagger\rho L_{i\alpha}-\frac{1}{2}\{L_{i\alpha}L_{i\alpha}^\dagger,\rho\}\right),
\label{eq:GAMEChain}
\end{equation}
with the renormalized Hamiltonian,
\begin{equation}
H'=H_0-\frac{i}{2}\sum_{i\alpha}(S_{i\alpha}S_{i\alpha,f}^\dagger-S_{i\alpha,f}S_{i\alpha}).
\end{equation}
The Lindblad generators are obtained from Eq.~\ref{eq:ansatz}, e.g. $L_{i\alpha}=S_{i\alpha}\circ \sqrt{\gamma}$.
We solve these equations by numeric iterations in the truncated Hilbert space, as explained in appendix~\ref{sec:App1}.

\subsection{\label{sec:UProcesses} Nearly Unitary Magnetization Dynamics}
\begin{figure}
\centering
\includegraphics[width=.8\textwidth]{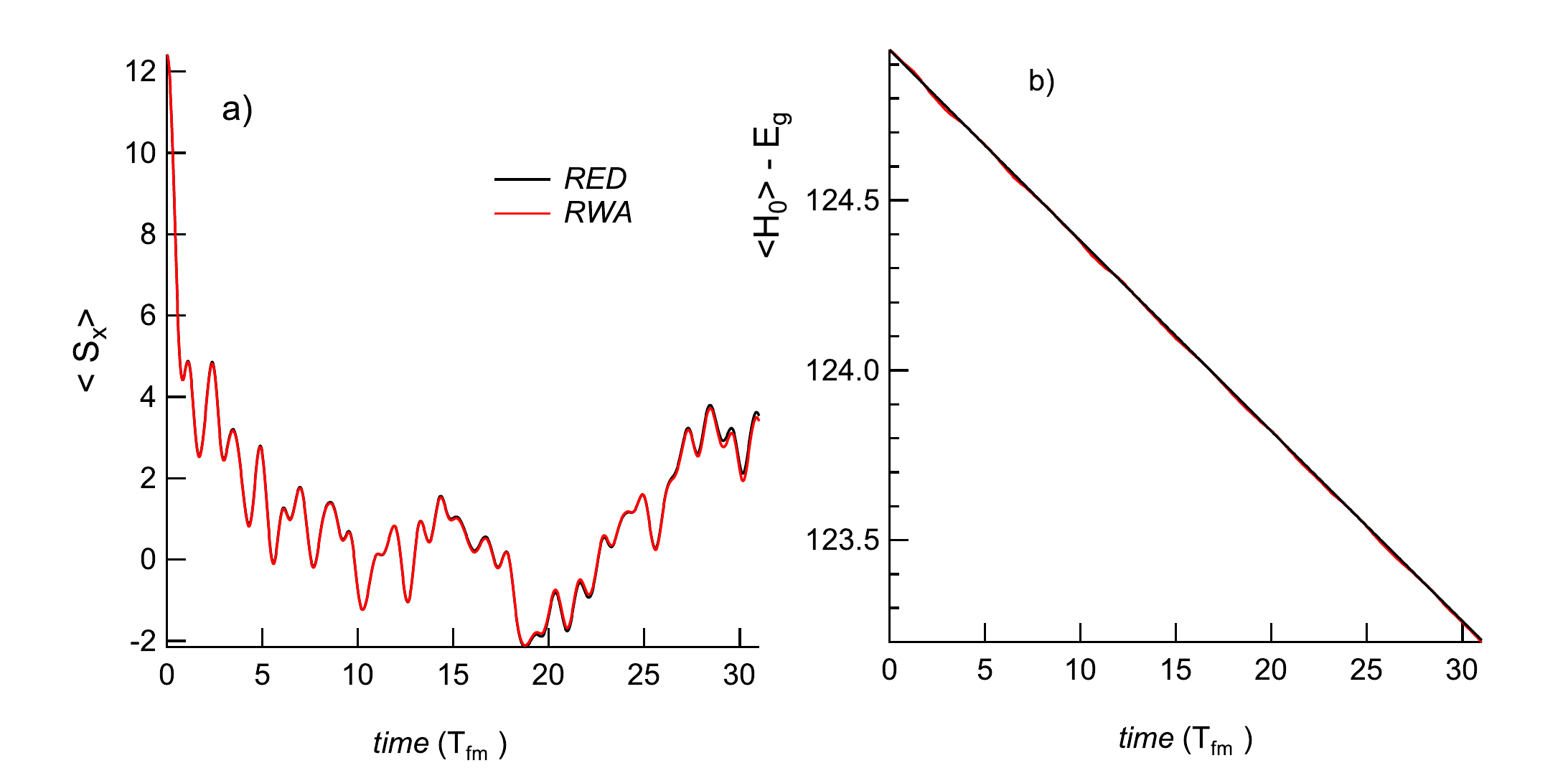}
\caption{Almost unitary magnetization dynamics of the spin-chain in the regime of weak coupling to the bath. a) Rapid decay of the initially perpendicular magnetization due to
the effective phase randomization of the many-body states of the spin-chain. b) Energy of the spin-chain versus time. $n=25$, $N=1728$, $g_{tot}=0.01$, $T_{fm}=2\pi/\Delta$, $\Delta=20.1$. \label{fig:unitary}}
\end{figure}

Let us first familiarize with the system by considering what we call very weak coupling to the heat bath,
e.g., we assume $g=0.000133$ for the SD in each bath.
Since there are 75 independent baths, the total coupling is $g_{tot}=75g=0.01$.

Fig.~\ref{fig:unitary}(a) displays a rapid decay of the perpendicular magnetization.
The bottom axis is time in units of
$T_{fm}$, where $T_{fm}=2\pi/\Delta$ is the period of the ferromagnetic resonance.
We consider the dynamics as mostly unitary, in a sense that $Tr \rho^2$ changes from one at $t=0$ to $0.744$ after thirty FMR-periods.
The decomposition of the initially uniformly magnetized state into
eigenstates, displayed in Fig.~\ref{fig:spectrum}(b), results in effective loss of magnetic coherence over the period of the FMR.
Over that time, $Tr(\rho^2)$ drops by only $0.01$, so the system retains quantum coherence even when the magnetization drops close to
zero.

The energy of the system $\langle H_0\rangle$ is shown in Fig.~\ref{fig:unitary}(b). It decays slowly due to the dissipative coupling to the heat bath.
The small drop of  $\langle H_0\rangle$, relative to the initial energy,
indicates that the figure displays an early stage of the relaxation cascade.

\subsection{\label{sec:RProcesses}Dissipation Time Scales}

\begin{figure}
  \centering
    \includegraphics[width=0.99\textwidth]{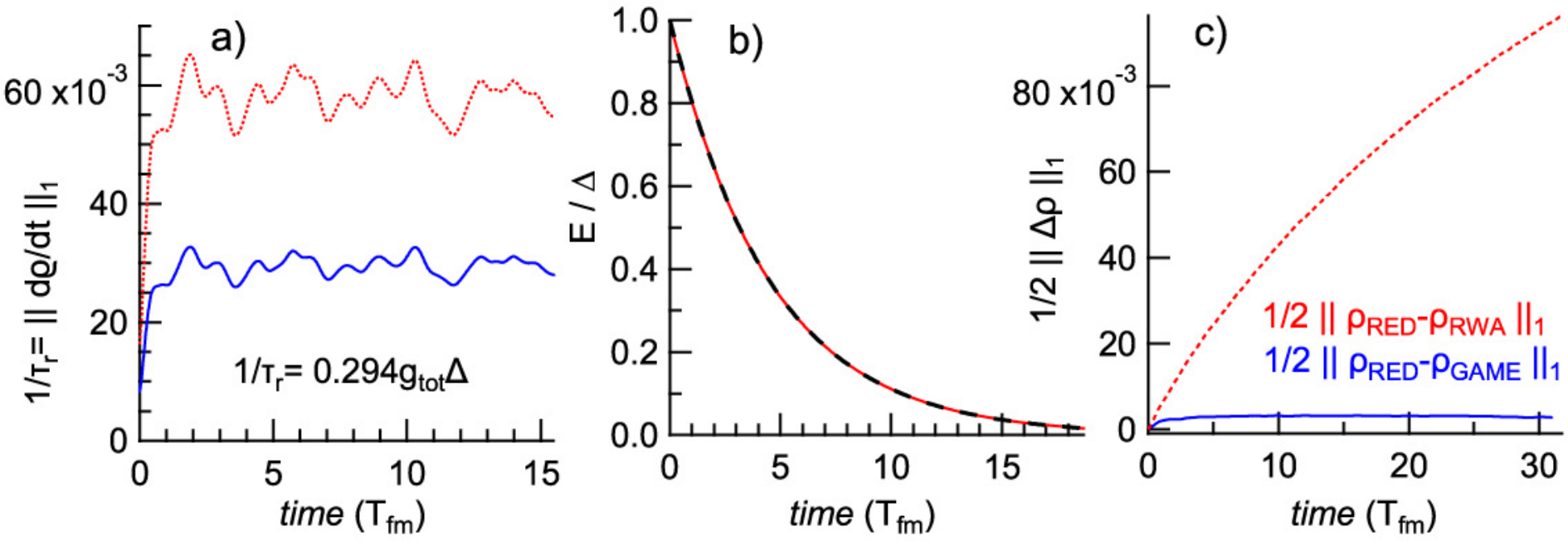}
  \caption{  a) State relaxation rate versus time, for $g_{tot}=0.005$ (blue-solid) and $0.01$ (red-dotted). b) Energy relaxation of the FMR at $g_{tot}=1$. Dashed thick-black line is the best fit to $\exp(-\Gamma't)$.  $\Gamma'=0.220T_{fm}^{-1}=0.035\Delta$ is the FMR width. c) Trace-distance between the solutions of the Redfield equation
and the approximate states in the RWA (red-dotted) and GAME (blue-solid), $g_{tot}=0.01$. In all panels, $n=25$, $N=1728$, $\omega_c=6\Delta$, $T_{fm}=2\pi/\Delta$, $\Delta=20.1$.\label{fig:relax1}}
\end{figure}

Now we examine the time-scale on which the environment changes the system properties in the interaction picture.
First we evaluate the density matrix relaxation rate $1/\tau_r$ for two initial conditions: a uniform perpendicular magnetization and the
first excited state.

To determine $\tau_r$,
we rotate into the interaction picture $\varrho=\exp(iH_0t)\rho\exp(-iH_0t)$, and determine $\tau_r$ as this,
\begin{equation}
\frac{\lvert\lvert\varrho_{RED}\rvert\rvert_1}{\tau_r}=\bigg|\bigg|\frac{d\varrho_{RED}}{dt}\bigg|\bigg|_1,
\end{equation}
where the subscript $RED$ indicates the state is obtained by solving the Redfield equation.  From now, we approximate this as
\begin{equation}
\frac{1}{\tau_r}=\bigg|\bigg|\frac{d\varrho_{RED}}{dt}\bigg|\bigg|_1
\label{eq:oneovertaur}
\end{equation}
since $\lvert\lvert\varrho_{RED}\rvert\rvert_1$  is only slightly larger than one. (The trace norm of the state can be larger than one if there are negative eigenvalues, but the effect is usually small in the range of applicability of the Redfield equation.)

Fig.~\ref{fig:relax1}(a) displays $1/\tau_r$ versus time, for the initially perpendicular magnetization, for $g_{tot}=0.005$ and $0.01$. The state relaxation rate $1/\tau_r$ scales with $g_{tot}$ as expected.
We find the rate $1/\tau_r$ by time-averaging, which calibrates the rate according to $1/\tau_{r}=0.293 g_{tot}\Delta$.
We similarly find that for the system initially prepared in the first excited state,
the state relaxation rate is $1/\tau_{E}=0.08g_{tot}\Delta$.

In many cases in micromagnetics, the parameter of interest is the energy relaxation rate, rather than the density-matrix relaxation rate.
To give a reference point to where our spin-chain system lies in that context, we find the energy relaxation rate of the first excited state of the system, e.g.,
the width of the ferromagnetic resonance. We calculate the system energy versus time, assuming the initial condition to be the first excited state. We fit energy versus time to $\exp (-\Gamma t)$,
where $\Gamma=1/\tau_E$ is the width we seek.
The one parameter fit is shown in Fig.~\ref{fig:relax1}(b), leading to the FMR width $1/\tau_{E}=0.0348 g_{tot}\Delta$. The
FMR is well resolved at our highest value of $g_{tot}=1$. In fact, the FMR width in that case corresponds to the Gilbert-damping parameter of $1/(4\tau_E \Delta)=0.0087$, which is not uncommon in transition metal ferromagnets.~\cite{Gilmore}

\begin{figure}
\includegraphics[width=.99\textwidth]{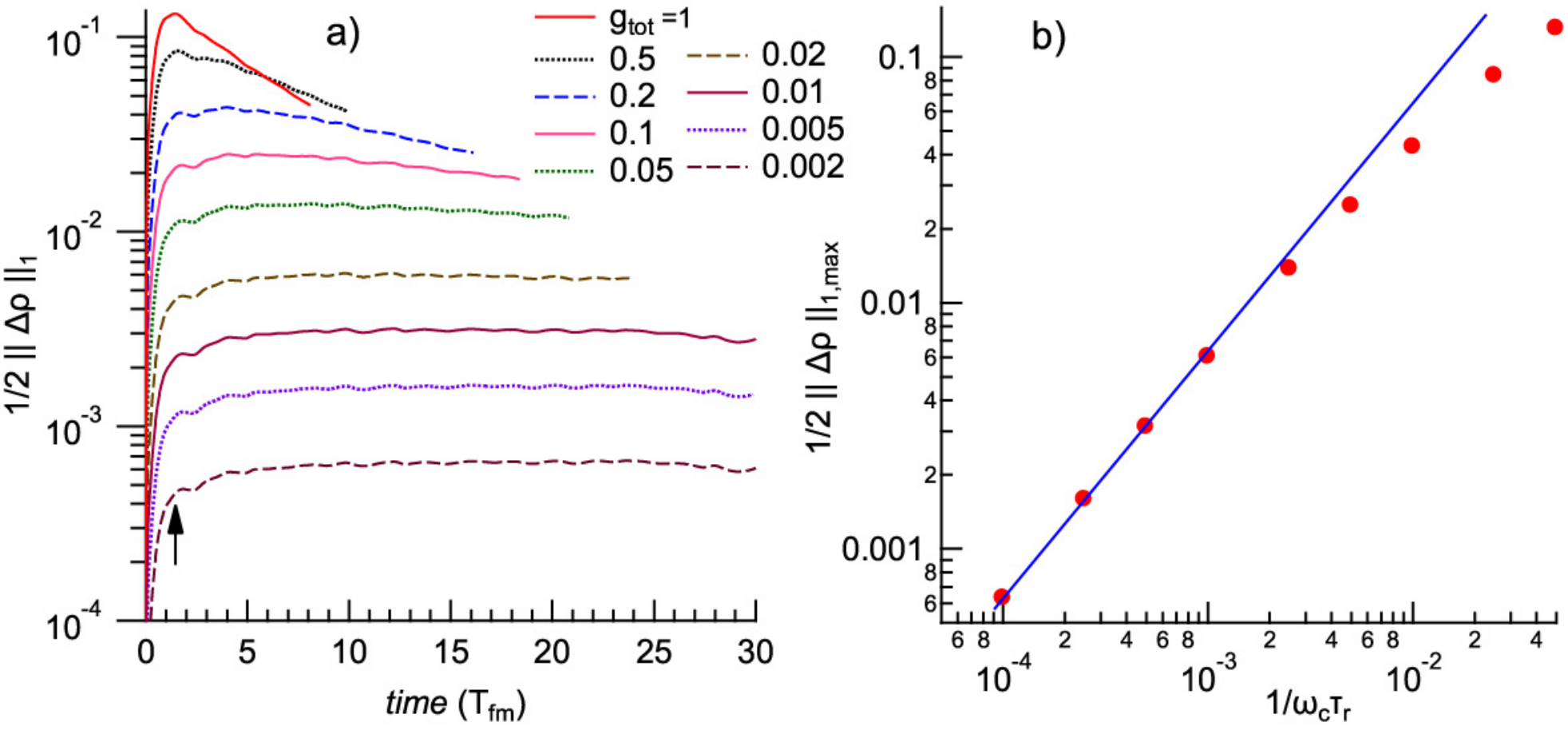}
\caption{ a) and b) display
the trace-distance between the solutions of the Redfield and GAME equations versus time, in the units of
the period of ferromagnetic resonance. $g_{tot}$ is the total
dimensionless coupling constant between the spin-chain and the heat bath.
b) Maximum trace-distance between the solutions of the two equations, versus rate $1/\tau_r$ in units of the cutoff frequency $\omega_c$.
The blue line is the linear scaling discussed in the text. $n=25$, $N=1728$, $\omega_c=6\Delta$, $T_{fm}=2\pi/\Delta$, $\Delta=20.1$.\label{fig:tauR}}
\end{figure}

In Fig.~\ref{fig:relax1}(c), we plot the trace-distance between the Redfield state and those of GAME and RWA, at $g_{tot}=0.01$. GAME clearly produces a state much closer to the Redfield state relative to that of the RWA.
In this weak coupling regime $1/\tau_{r}=0.00293\Delta$, and
the Born-Markov condition is well satisfied,
$1/\omega_c\tau_r\approx 5\times 10^{-4}$, assuring high accuracy of the Redfield equation.
Still, the RWA has poor accuracy since the smallest level spacing is much smaller than $1/\tau_{r}$.

We note in passing on the ability of trace-distance to distinguish quantum states. Fig.~\ref{fig:unitary} would naively suggest that the RWA accounts for the dynamical properties of the system spectacularly well, yet Fig.~\ref{fig:relax1}(c) tells a very different story, that the trace-distance between the RWA and the Redfield state in the same situation increases with time and is far from saturation.

To study the scaling of the trace-distance with system-bath coupling, in Fig.~\ref{fig:tauR}(a) we display the trace-distance $\frac{1}{2}\vert\vert\varrho_{RED}(t)-\varrho_{GAME}(t)\vert\vert_1$
versus time and the coupling strength $g_{tot}$. As the coupling decreases, we see that the trace distance versus time saturates on a time scale {\it independent} of the coupling $g_{tot}$. This introduces a new time scale of the system, the saturation time $T_s$. It is indicated by the arrow in Fig.~\ref{fig:tauR}(a), showing that it is comparable to $T_{fm}$.
In appendix~\ref{sec:Invariance} we show  that increasing the bath cutoff frequency above $6\Delta$  has a weak effect on $T_s$.

In Fig.~\ref{fig:tauR}(a) and at low $g_{tot}$, the trace-distance versus time saturates because of the transient nature of the detuning term that we drop from the Redfield equation. That is, the integral of $\varrho_{RED}(t)-\varrho_{GAME}(t)$ will saturate on time-scale over which the detuning superoperator averages out to zero. Thus
restoration of complete positivity can be viewed as a state slip on time scale $T_c$, with no further trace-distance increase after that time. A different state-slip at the correlation time scale of the bath is a known phenomenon in stochastic quantum dynamics,~\cite{Haake,Gaspard,Cheng,Suarez,TingPM} but here the slippage to regain complete positivity takes a longer, system dependent time.
The trace-distance accumulated over time $T_c$ is $T_c/\tau_r=gT_c/\tau_c$. It is linear with $g$, since neither $T_c$ nor $\tau_c$ depend on $g$.

In Fig.~\ref{fig:tauR}(b) we show the maximum trace-distance between the solutions of the Redfield and GAME equations versus state relaxation rate $1/\tau_r$.
 The figure displays asymptotic linear scaling between the trace-distance and the state relaxation rate at low $1/\tau_r$. The best linear fit on the lowest four points is
\begin{equation}
\frac{1}{2}\lvert\lvert \varrho_{RED}-\varrho_{GAME}\rvert\rvert_{1,max} \approx 6.41\frac{1}{\omega_c\tau_r}= \frac{1.07}{\Delta\tau_r},
\end{equation}
and is shown by the blue line.

\section{~\label{sec:benchmark}Comparisons Between Various Master Equations}.

There have been several recently derived CP master equations with the claim that they perform well in the limit of small level spacings of the quantum system. In addition, there are master equations with TDCs, that we discussed in Secs.~\ref{sec:RMA} and~\ref{sec:game} that have fewer problems than the Redfield equation with asymptotic coefficients.
In this section we evaluate
how much the solutions of different master equations can diverge from each other for the same initial condition.

For the initial condition we use the same uniformly polarized state perpendicular to chain axis.
In Sec.~\ref{sec:TDC} we also study the decay of the FMR.
For the coupling to the Ohmic bath, we use $g=0.0133$ per bath ($g_{tot}=1$ for the 25-site chain), the highest value in Fig.~\ref{fig:tauR}, where the break-down of the master equation is hinting.
Also we study the case $g=0.000133$ per bath ($g_{tot}=0.01$ for the 25-site chain), which we refer to in  Sec.~\ref{sec:UProcesses} as "nearly unitary" dynamics.
While the higher value of $g_{tot}$ is more realistic in terms of damping in magnetic materials, the Redfield approach is perhaps on shaky grounds. For the smaller $g_{tot}$, the Redfield equation is valid.

In all examples we use the same exponential cutoff with frequency of $\omega_c=6\Delta=120$. In this case $\tau_c/\tau_r=0.0493$ for the larger $g_{tot}$ at $n=25$,
so we are comfortably but perhaps not too comfortably within the limit of the validity of the Bohr-Markov approximation.
Since different master equation have different complexities, we adjust the code for fastest performance, depending on the approximation we use. The details will be provided as we go through the approximations.

\subsection{\label{sec:PRWA} Partial Rotating Wave Approximation.}

The partial RWA (PRWA) extends the range of applicability of the RWA.~\cite{Vogt,Tscherbul,Jeske,Hartmann} It begins by observing that if the SD is flat, then the Redfield equation will
be in Lindblad form. The approximation works as follows.
First, the system Bohr energies or frequencies $\omega_{nm}$ are sorted and divided into bins.

In the next step, the Redfield equation (Eq.~\ref{eq:RedDD}) is truncated, but less draconically so than in the RWA. In the first line of Eq.~\ref{eq:RedDD}, for example, all terms except those where $\omega_{jm}$ and $\omega_{in}$ belong to the same bin are discarded. At the same time, the SD within the bin is approximated by an average (and therefore flat) SD, one for each bin. As a result, each bin contributes to a Lindblad generator specific to that bin, and different bins are decoupled. In our case we find the average frequency within each bin, (we call it the coarse-grained frequency), and calculate the SD at that frequency. We checked that two other frequency averaging techniques give approximately the same results.
This binning is also performed on the loss terms in Eq.~\ref{eq:RedDD}.

\begin{figure}
\centering
\includegraphics[width=\textwidth]{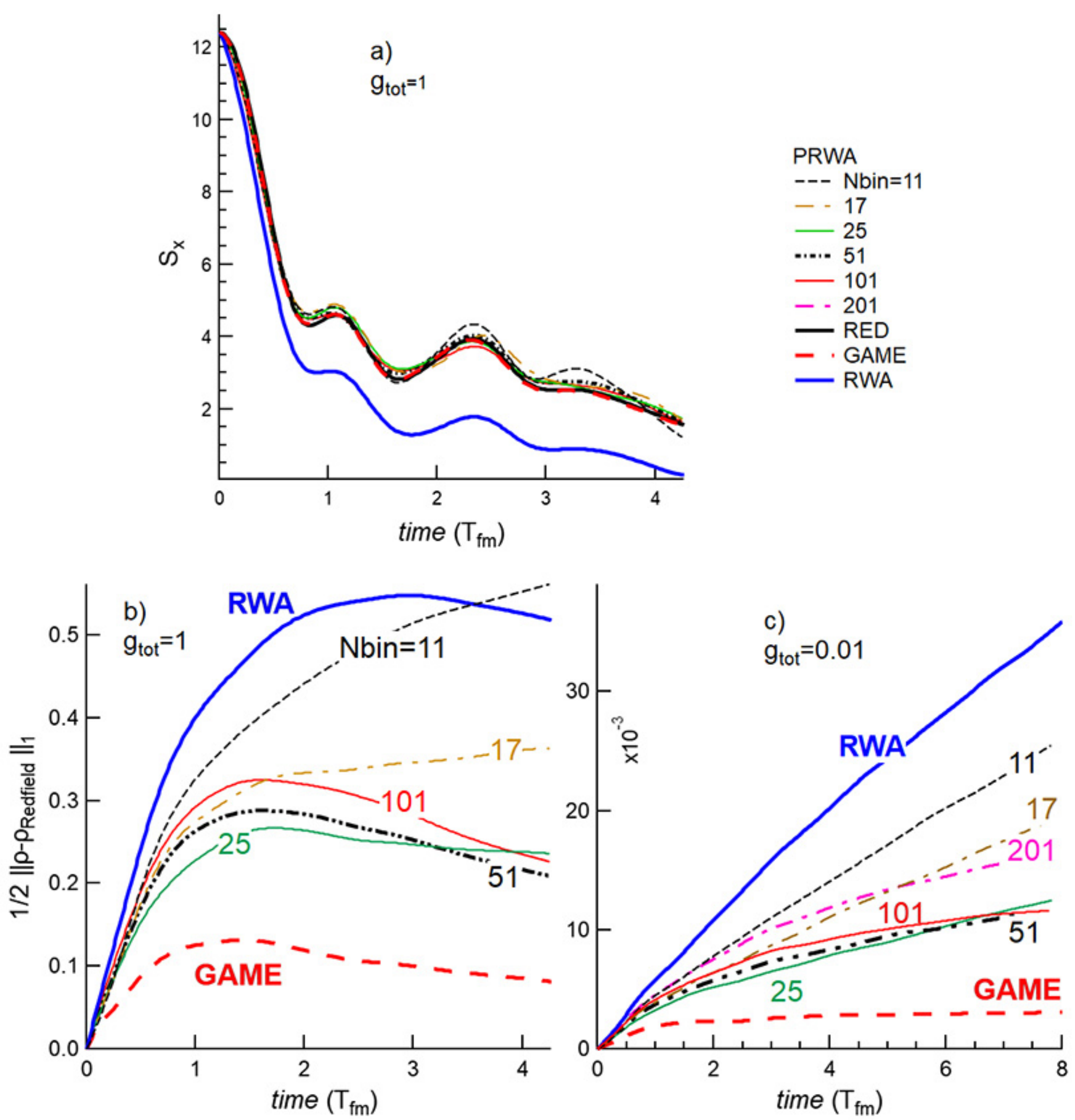}
\caption{ Partial Rotating Wave Approximation. a) Magnetization of the 25-site spin chain versus time, in units of the ferromagnetic resonance period. The Redfield and GAME solutions are virtually the same. $g_{tot}=1$. b) Trace norm distance to the solution of the Redfield equation versus time, from the states solved using the RWA, GAME, and PRWA.
$g_{tot}=1$. c) The same as b), but with $g_{tot}=0.01$. The trace-distance in c) is far from saturation in all approximations except GAME. $N=1728$, $\Delta = 20$, $\omega_c=6\Delta$. Ohmic bath with exponential frequency cutoff.
\label{fig:PRWA}}
\end{figure}

The bin width ($\delta b$) is an adjustable parameter, but the least compromising
value seems to be set by the relaxation rate $\tau_r^{-1}$. If $\delta b\ll 1/\tau_r$, then the PRWA will have the same problem as the RWA: the oscillations at the coarse-grained frequency differences will not average out
on time scale $\tau_r$, resulting in poor accuracy. If on the other hand $\delta b\gg 1/\tau_r$, the assumption of the flat SD becomes less valid.

The number of Lindblad generators increases with decreasing bin width. The RWA is the limit of the PRWA when there is one frequency per bin. Depending on the bin width, it may take significant resources to numerically perform the binning and obtain the generators. For example, it takes approximately four hours to count the frequency duplicates and find the generators in the RWA, for the 25-site Heisenberg chain with $N=1728$.
Fortunately, the sparseness of the generators also increases with the decreasing bin width, so the issue of time and memory is managed using sparse matrices.

Note that for the $n=25$ spin-chain, the relaxation rate of the state in the interaction picture
$1/\tau_r=5.91$ at $g_{tot}=1$, while the frequency span is $\approx 316$.
So we expect that the optimum number of bins of $316/5.91\approx 53$ for $g_{tot}=1$. In that case the average number of system frequencies per bin is approximately $6\times 10^4$,  and it fluctuates widely between the bins. In the case of $g_{tot}=0.01$, the optimum bin width should be much lower. But
we do not solve the PRWA equation over time scale $\tau_r$, because it would take unreasonable simulation time.

The renormalization of the system Hamiltonian given by Eq.~\ref{eq:LambDD} in a sense completes the PRWA, as follows. Eq.~\ref{eq:RedDD}
shows that flat spectral functions result in the cancellation of the PDs on the RHSs. Had we not renormalized the Hamiltonian according to Eq.~\ref{eq:LambDD},
the principal density would not cancel in the loss term. The renormalization therefore completes the PRWA by accounting the PDs via the renormalized Hamiltonian, thereby insuring correctness.
Previous approaches involved either discarding the Lamb shift,~\cite{Vogt} or coarsegraining the RWA Lamb-shift to the representative bin frequency.~\cite{Hartmann} The renormalized Hamiltonian in Eq.~\ref{eq:shif} does not need any binning so we apply it here in that primitive form.

In Fig.~\ref{fig:PRWA} we vary the number of bins between $11$ and $201$  and find the corresponding solutions in the PRWA.  In Fig.~\ref{fig:PRWA}(a) numerical calculations show that the overall magnetic dynamics is well accounted for by the PRWA, especially when compared to the
RWA.

The traces distance between the solutions of the RWA, PRWA, GAME and the Redfield equation is shown in Fig.~\ref{fig:PRWA}(b,c).
Initially all trace-distances are zero because the states are the same. As time increases, the states under different master equations evolve differently and the trace distances increase. At long times, however, as the states relax towards the ground state the trace-distances become suppressed. Thus, a trace-distance versus time exhibits a maximum.

The optimum bin-width corresponds to the trace-distance with the smallest maximum.
For $g_{tot}=1$,  the optimum bin width is in the expected range.
At weak coupling, ($g_{tot}=0.01$), the trace-distances to the PRWA solutions versus time do not saturate,
while the trace-distance between GAME and the Redfield equation saturates at time similar to the FMR-period.

In summary, we find that GAME is a significantly more accurate approximation of the Redfield equation than the PRWA. Since both GAME and PRWA are derived by approximating the same Redfield equation, the former is most likely also a significantly better approximation of true reduced quantum dynamics.

\subsection{\label{sec:DCGA} Dynamical Coarse-Graining Approximation}

As an example of completely positive coarse-grained  master equations, here we study the properties of the dynamically coarse-grained (DCG) master equation.~\cite{Schaller,Benatti}
The equation is derived by coarse-graining of the Liouville equation. The derivation applies the Born, but not Markov approximation.
In the vectorized from of the density matrix, the DCG equation utilizes the Liouville superoperator, which is provided
explicitly by Schaller and Brandes.~\cite{Schaller}
After translating to the notation we use in this article, we find that the DCG-master equation reads the same as the Redfield equation in the form of Eq.~\ref{eq:GRED},
except that the dissipative kernel $G(\omega,\omega')$ in Eq.~\ref{eq:Gtensor} is replaced with
\begin{equation}
G_\tau^{dc}(\omega,\omega')=\frac{\tau}{2\pi} e^{-i\frac{(\omega-\omega')\tau}{2}}\int_{-\infty}^{\infty}\gamma(\Omega)\,\text{sinc}\frac{(\Omega-\omega)\tau}{2}\,
\text{sinc}\frac{(\Omega-\omega')\tau}{2}\,d\Omega,
\label{eq:ddc}
\end{equation}
where $\tau$ is the coarse-graining time. Note that our double-indices in Eq.~\ref{eq:Gtensor} are different from Ref.~\cite{Schaller}, i.e., $G_{dc,ab}(\tau)=\gamma_{cd,ba}(\tau)$, where
$\gamma_{cd,ba}(\tau)$ is the dissipative kernel given by Eq.~23 in that reference.

As alluded to by Schaller and Brandes,~\cite{Schaller} this expression can be decomposed in terms of single-variable functions only,
\begin{equation}
G_\tau^{dc}(\omega,\omega')=e^{-i\frac{(\omega-\omega')\tau}{2}}\left\{
\frac{1}{2}[\gamma_\tau(\omega')+\gamma_\tau(\omega)]\text{sinc}\frac{(\omega'-\omega)\tau}{2}-
\frac{S_\tau(\omega')-S_\tau(\omega)}{(\omega'-\omega)\frac{\tau}{2}}\cos\frac{(\omega'-\omega)\tau}{2}
\label{eq:kernelDC}
\right\}
\end{equation}
which is critical to our numerical calculation. Namely, to determine the $N^4$ values of $G_\tau^{dc}(\omega,\omega')$,
it suffices to calculate the functions $\gamma_\tau(\omega)$, $S_\tau(\omega)$, and $\partial S_\tau(\omega)/\partial\omega$, thereby reducing the computational complexity from $O(N^4)$ to $O(N^2)$. We show in appendix~\ref{sec:TDEPSF} that $\gamma_\tau(\omega)$ are $S_\tau(\omega)$ are the time-dependent spectral and principal densities, respectively, as defined by Eq.~\ref{eq:GammaTDC}. The kernel $G_\tau^{dc}(\omega,\omega')$ is a positive semidefinite $N^2\times N^2$ matrix,
which assures that the DCG equation is in a Lindblad form.~\cite{Schaller}

In addition, the renormalized Hamiltonian $H$ in Eq.~\ref{eq:LambDD} is replaced with
\begin{equation}
H_{\tau,nm}^{dc}=E_n\delta_{nm}+\sum_{i}{\mathcal H}_\tau^{dc}(\omega_{ni},\omega_{mi})A_{ni}A_{im},
\label{eq:dcgLAa}
\end{equation}
with the unitary kernel
\begin{equation}
{\mathcal H}_\tau^{dc}(\omega,\omega')=\frac{\tau}{2\pi}e^{i(\omega-\omega')\tau/2}
\int_{-\infty}^{\infty}S(\Omega)\,\text{sinc}\frac{(\Omega-\omega)\tau}{2}\,\text{sinc}
\frac{(\Omega-\omega')\tau}{2}\,d\Omega.
\label{eq:dcgLAb}
\end{equation}
We decompose the unitary kernel similar to the dissipative one, as described in more detail in the appendix,
which yields the explicit formula for the kernel that governs the unitary contribution of the heat bath to system quantum dynamics:
\begin{equation}
{\mathcal H}_\tau^{dc}(\omega,\omega')=e^{i\frac{(\omega-\omega')\tau}{2}}\left\{
\frac{1}{2}[S_\tau(\omega')+S_\tau(\omega)]\text{sinc}\frac{(\omega'-\omega)\tau}{2}+
\frac{1}{4}\frac{\gamma_\tau(\omega')-\gamma_\tau(\omega)}{(\omega'-\omega)\frac{\tau}{2}}\cos\frac{(\omega'-\omega)\tau}{2}
\right\}.
\label{eq:dcReactance}
\end{equation}

\begin{figure}
\centering
\includegraphics[width=0.89\textwidth]{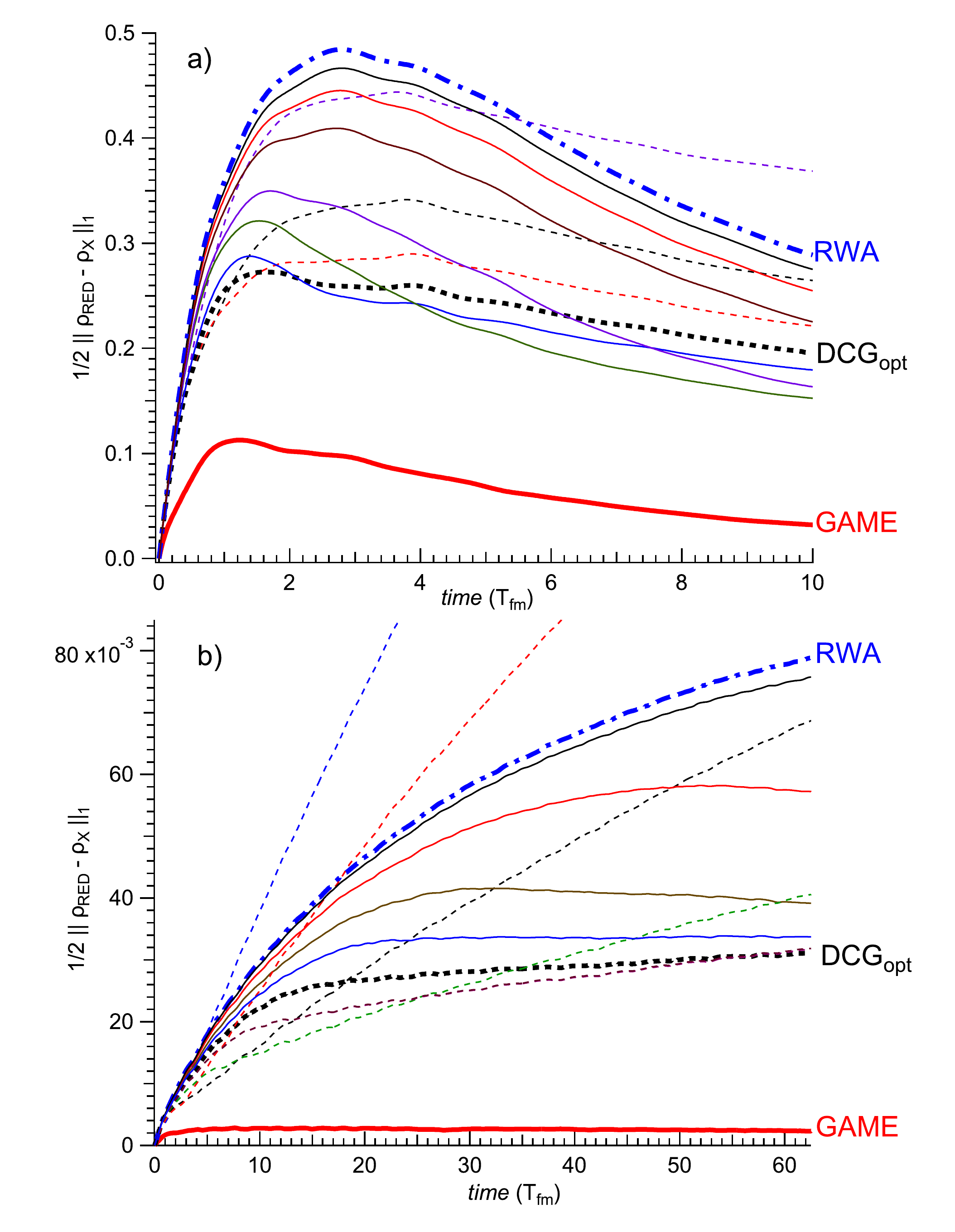}
\caption{Trace-distances between the
solutions of various master equations and that of the Redfield equation, versus time. a) Regime of moderate environmental coupling $g_{tot}=69g=0.92$. Thick lines: RWA  (dashed-blue), $\text{DCG}_\text{opt}$ (dashed-black), and GAME (red-full). The approximate optimum coarsegraining time is $\tau_{opt}=1.3$. Thin-full lines: coarse-graining times $\tau=64,32,16,6.4,3.2$, and $1.6$, follow the maxima between the RWA and $\text{DCG}_\text{opt}$ from top to bottom, respectively.
Thin-dashed lines: $\tau=0.96,0.64,0.32,$ and $0.16$, follow the maxima from $\text{DCG}_\text{opt}$, bottom to top, respectively.
b)  Analogous to a) in the weak environmental coupling regime $g_{tot}=0.0092$. $\tau_{opt}= 16$. Thin-full lines, top-to-bottom: $\tau=160,64,32,$ and $22.4$. Thin-dashed lines, bottom-to-top:
$\tau=11.2,6.4,3.2,1.6,$ and $0.96$.
All time scales are in units of $T_{fm}$.
Ohmic SD with exponential cutoff, $n=23$, $N=694$, $\Delta=20.1$, and $\omega_c=6\Delta$.
\label{fig:DCGA}}
\end{figure}

Here we utilize a 23-site spin-chain and solve the DCG-master equation at various coarse-graining times, assuming a perpendicularly magnetized state as the initial condition.
In Fig~\ref{fig:DCGA}, we display the trace-distances from the solution of the Redfield equation,
for Ohmic SD with exponential frequency cutoff.
The trace-distances versus time
exhibit maxima as discussed in the previous section. The curve with the smallest maximum defines the optimum coarse graining time. For the weak environmental coupling [Fig~\ref{fig:DCGA}(b)], it takes unreasonable amount of time to calculate how the curves merge at $t\to\infty$
and the maximum may not be reached within the time range of the simulation. In that case we apply the maximum trace-distance within the range of the figure.
As $\tau$ decreases from $\infty$ to $\tau_{opt}$, the maximum trace-distance of the DCG-state decreases. However, below $\tau_{opt}$
the maximum trace-distance increases with decreasing coarse-graining time. [At $\tau=0$, the trace-distance varies as $1/2\,\vert\vert\varrho_{RED}(t)-\rho_0\vert\vert_1$]. At $g_{tot}=0.92$ and $0.0092$, $\tau_{opt}=1.3T_{fm}$ and $16T_{fm}$, respectively, in good agreement with square root dependence on $g$, which makes it consistent with the coarse-graining approximation.~\cite{Majenz,mozgunov} The corresponding values of $\sqrt{\tau_c\tau_r}$ (the optimum coarse-graining time in the CGSE, see Sec.~\ref{sec:TDC}) are approximately factor of 10 smaller than the values we find.

It is striking how in Fig~\ref{fig:DCGA}(b) the trace-distance of GAME saturates at a time scale $\tau_s$ much smaller than in any other approximation.
This results in improved accuracy of GAME, assuming that the figure of merit is the trace distance to the Redfield state.
The rapid saturation of the trace-distance (also seen in Fig.~\ref{fig:PRWA}) is a specific trait of GAME. It is due to the transient nature of the term that we dropped from the Redfield equation in order to restore CP.
The saturation is less pronounced for moderately weak environmental coupling [Fig~\ref{fig:DCGA}(a)], due to the more rapid relaxation which prevents us
from reaching the flat part of the trace-distance curve.

We also perform numerical calculations assuming Ohmic SD with the Drude-Lorentz frequency cutoff. In the Drude-Lorentz case we obtain the coefficients of the DCG master equation using a different numerical method, yet we find that results are essentially the same, e.g., see Fig.~\ref{fig:DCGB} in appendix~\ref{sec:TDEPSF}.

To summarize, we again find that GAME is a significantly more accurate approximation of the Redfield equation.  However, in contrast to the previous section, now the compared approximations operate under different premises, because they are differently derived from first principles. Here the improved  approximation of the Redfield state does not necessarily imply that GAME is an improved approximation of the true reduced quantum dynamics. Rather, it opens a question of why there is such a difference between the Redfield and DCG approximations in the first place. Since the focus of this paper is to find a simple and accurate CP-approximation of the Redfield equation, we do not speculate on possible answers to this question at this time.
\pagebreak
\subsection{\label{sec:PERLind} $\sqrt{\text{SD}}$-Approximations}

The Lamb-shifts being the only difference between various $\sqrt{\text{SD}}$-approximations, let us investigate them in detail.  The PERLind approach does not utilize a Lamb-shift.~\cite{perlind} Out of curiosity, we
add to it the Lamb-shift from the RWA to see if it improves the accuracy.
ULE, on the other hand, does provide the Lamb-shift~\cite{Nathan}, that translates to our notation as
\begin{equation}
H_{nm}=E_n\delta_{nm}+\sum_{i}H^{ule}(\omega_{ni},\omega_{mi})A_{ni}A_{im},
\label{eq:Lund}
\end{equation}
with the  kernel given by
\begin{equation}
{\mathcal H}^{ule}(\omega,\omega')=-\frac{1}{2\pi}\mathcal{P}\int_{-\infty}^{\infty} \frac{d\Omega}{\Omega}\sqrt{\gamma(\Omega+\omega)\gamma(\Omega+\omega')}.
\label{eq:Lund1}
\end{equation}
The diagonals of the RWA, GAME, and ULE Lamb-shifts are identical. (In ULE the diagonal calculation involves a  Kramers-Kronig
transform given in table~\ref{table1}.) So the differences between various Lamb-shifted $\sqrt{\text{SD}}$-approximations are all in the off-diagonal matrix elements of the Lamb-shift.

In terms of numerics, the ULE Lamb-shift requires $N^3$ numerical integrations, as we could not find a way to decompose
the kernel in Eq.~\ref{eq:Lund1} into single-variable functions of frequency,
while GAME needs only $N^2$ analytical calculations of the spectral functions.
After optimizing the codes for ULE and GAME, in the example $n=25$ and $N=1728$ it takes approximately 61 hours to compute the Lamb-shift of ULE, compared to approximately 30 seconds of that in GAME.
Similarly, in terms of memory, ULE requires storage of a rank 3 tensor (Eq.~\ref{eq:Lund}), while GAME only stores 2D matrices of SDs and PDs at Bohr frequencies.

\begin{figure}
  \centering
    \includegraphics[width=\textwidth]{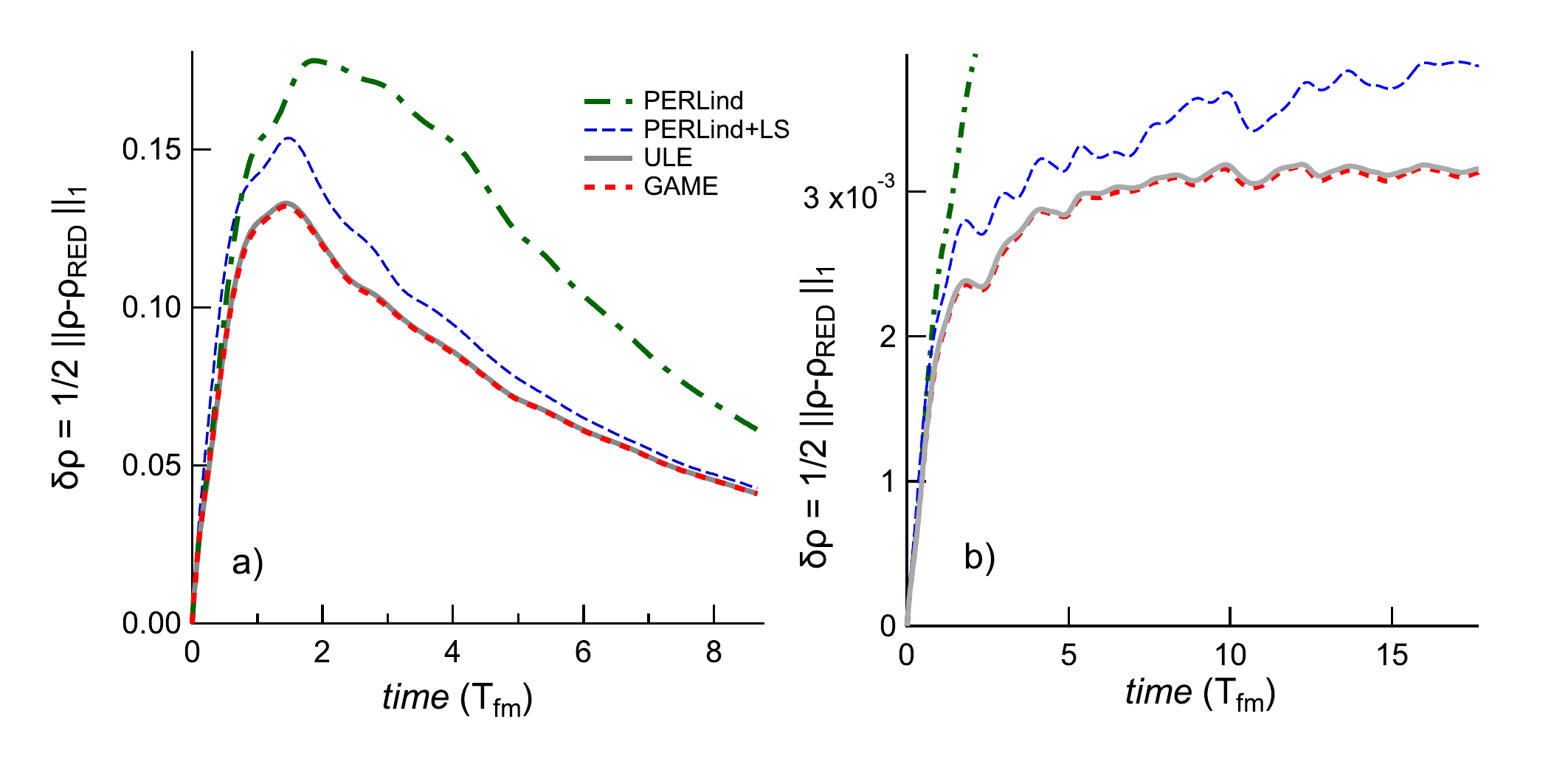}
  \caption{Comparison of four approximations: PERLind, PERLind with the added RWA Lamb-shift, ULE, and GAME. Trace norm distances are taken with respect to the Redfield equation solutions. a) $g_{tot}=1$ and b) $g_{tot}=0.01$. $n=25$, $N=1728$, $\omega_c=6\Delta$, $T_{fm}=2\pi/\Delta$, $\Delta=20.1$.
\label{fig:ansatz}}
\end{figure}
In Fig.~\ref{fig:ansatz} we present the trace-distance to the solution of the Redfield equation, from the solutions of the  PERLind, PERLind plus LS from the RWA, ULE, and GAME.
In Fig.~\ref{fig:ansatz}(a) the  coupling to the heat bath $g_{tot}=1$ is the highest one from Fig.~\ref{fig:tauR}, while in Fig.~\ref{fig:ansatz}(b) the coupling is $g_{tot}=0.01$

GAME and ULE are clearly the closest to the Redfield solutions. The Lamb-shifts of GAME and ULE are weakly off-diagonal in this example. Nevertheless, the small off-diagonal elements matter
significantly, since PERLind with the added RWA Lamb-shift (which is diagonal) has significantly higher error than both ULE and GAME.

GAME is slightly closer to the Redfield state than ULE, as can be clearly seen in Fig.~\ref{fig:ansatz}(b). This is not surprising because GAME uses the implicit Lamb-shift of the Redfield equation. It is nevertheless surprising how close the two approximations are,
even though the Lamb-shift calculations are quite different.

\subsection{\label{sec:TDC} Master Equations with Time Dependent Coefficients.}

\begin{wrapfigure}{R}{.6\textwidth}
\centering
\includegraphics[width=.59\textwidth]{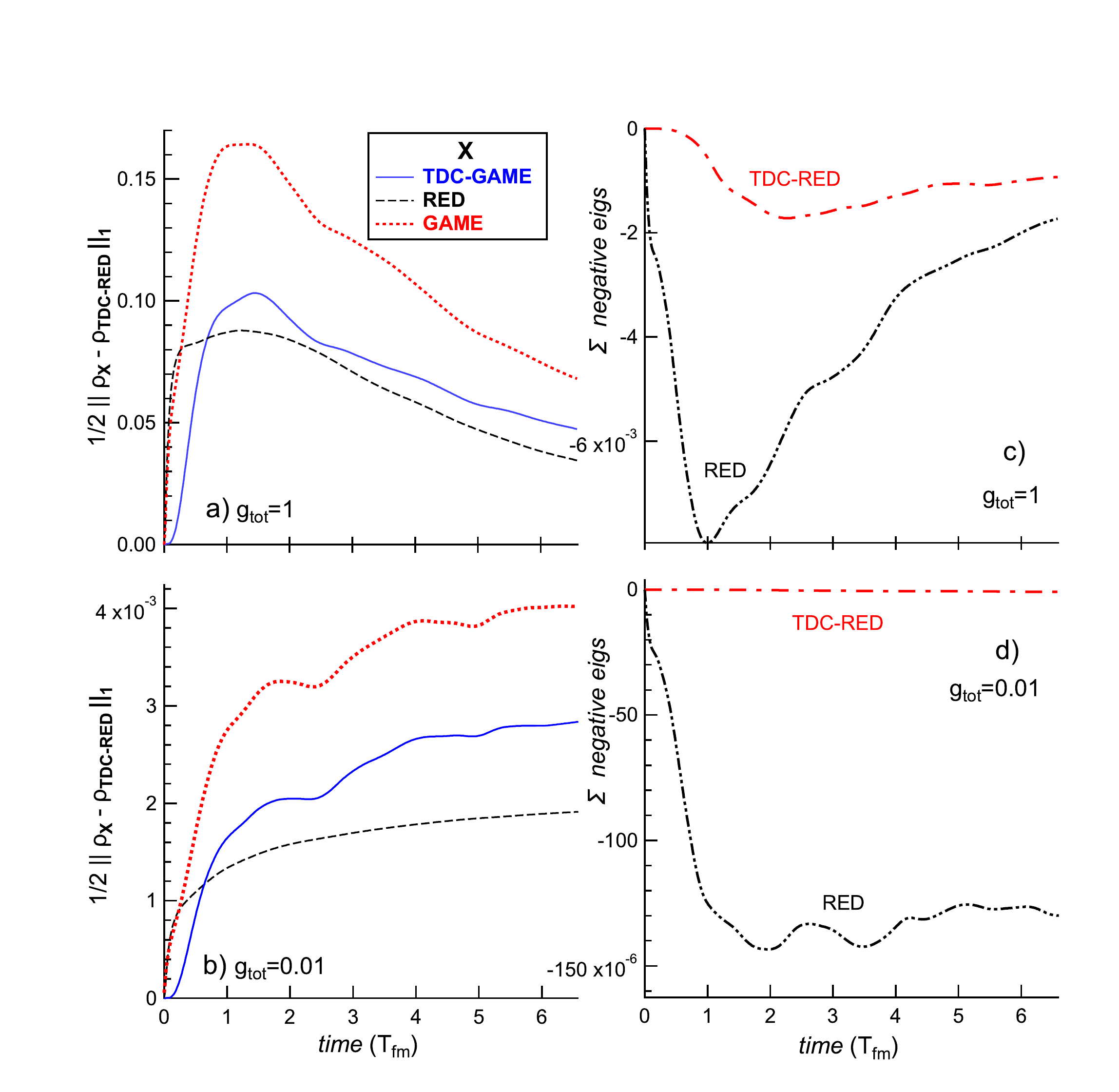}
\caption{Decay of the perpendicularly magnetized state. a) and b): Trace-distances between the solutions of  equations X and the TDC-Redfield equation, at $g_{tot}=1$ and $0.01$, respectively. There is a separation in time scales for divergence of the trace-distance in the Redfield versus TDC-GAME. The former diverges on a time scale of the bath, while the latter diverges on a time scale comparable to $T_{fm}$. c) and d): Sum of the negative eigenvalues of the TDC-Redfield and Redfield solutions versus time, for  $g_{tot}=1$ and $0.01$, respectively. $n=25$, $N=1728$, $\omega_c=6\Delta$, $T_{fm}=2\pi/\Delta$, $\Delta=20.1$.
\label{fig:TDC}}
\end{wrapfigure}

Here we compare the solutions of the Redfield, GAME, and TDC-GAME equations to that of the TDC-Redfield equation (recall diagram in Fig.~\ref{fig:norma}).
Since the TDC-Redfield equation is the least approximated among the four, it is likely the most accurate.

In Fig.~\ref{fig:TDC} we study the decay of the initially perpendicularly magnetized state. In particular, (a) and (b) show numerically obtained trace-distances versus time at $g_{tot}=1$ and $g_{tot}=0.01$, respectively. At early times, defined here as $t<T_{fm}$, the trace-distance between the TDC-Redfield and the Redfield state increases rapidly with a characteristic time scale much smaller than $T_{fm}$, independent of $g$.
In comparison, the trace-distance between the TDC-Redfield and TDC-GAME increases with a delay, on a
time scale comparable to the characteristic system time (FMR-period), also independent of $g$. This suggests there is a separation of time scales that govern the two approximations: one that restores CP and the other that replaces TDC-coefficients with their asymptotic values. We interpret the rapid initial increase in the trace-distance on time scale $\ll T_{fm}$ in terms of the initial state slip we mentioned earlier.~\cite{Haake,Gaspard,Cheng,Suarez,TingPM}

Note that now is not the first time that a CP-restoration is governed by a system-specific time-scale, that is independent of the system-bath coupling strength. The RWA is a well known example, where that time-scale is the Heisenberg time. In GAME CP restoration, the time scale is similar to the characteristic system time, in this case the FMR-period, which is much smaller than the Heisenberg time. This suggests that GAME may be most suitable for applications on quantum systems with a dominant system frequency, which is the case in ferromagnets or superconductors, or any other gapped system.

Other coarsegraining approximations such as CGSE and PRWA work quite differently. They usually involve some optimization process that depends on $g$. For example, the CGSE equation has optimum coarse-graining time of order $\sqrt{\tau_r\tau_c}$ but it seems that there is no system dependence of the optimum time. Thus CGSE
may be applicable even if $1/\tau_r$ is larger than the dominant system frequency, if $\omega_c$ is large enough. We note that in such regime the Redfield equation is unlikely to be stable. The question of which master equation works best in what regimes is a complicated matter and will remain the subject of future reasearch.

Fig.~\ref{fig:TDC}(c,d) presents the sum of the negative eigenvalues of the solutions of the Redfield equations, corresponding to $g_{tot}=1$ and $0.01$, respectively. In (c), both Redfield equations develop significant negativity. This would suggest that there is  a breakdown of the perturbative
approach, according to the analysis by Hartmann and Strunz.~\cite{Hartmann} It suggests that quantum theory of magnetics may need to extend beyond the master-equation platform and involve regime of strong coupling to the heat bath.
On the other hand, in Fig.~\ref{fig:TDC}(d), the TDC-Redfield equation has strongly suppressed negativity, which implies the validity of the weak coupling regime.

\begin{wrapfigure}{L}{0.6\textwidth}
\centering
\includegraphics[width=0.59\textwidth]{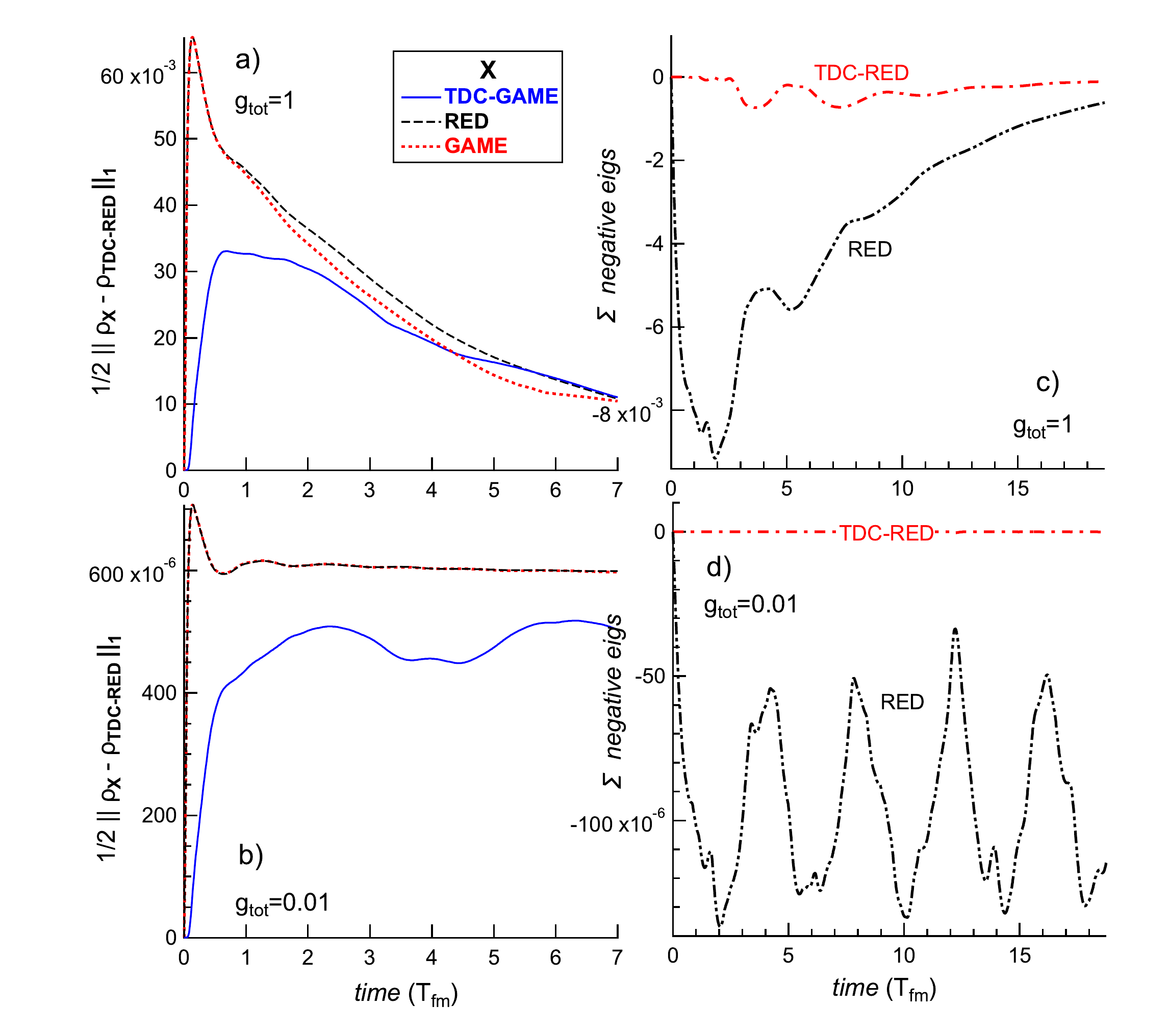}
\caption{Decay of the ferromagnetic resonance. a) and b): Trace-distances between the solutions of equations X and the TDC-Redfield equation, at $g_{tot}=1$ and $0.01$, respectively. c) and d): Sum of the negative eigenvalues of the TDC-Redfield and Redfield solutions versus time, for  $g_{tot}=1$ and $0.01$, respectively. $n=25$, $N=1728$, $\omega_c=6\Delta$, $T_{fm}=2\pi/\Delta$, $\Delta=20.1$.
\label{fig:TDCfmr}}
\end{wrapfigure}

We have investigated the properties displayed in Fig.~\ref{fig:TDC} for different initial conditions. In the specific example of the lowest excited state (FMR), the TDC-GAME is always closest to the TDC-Redfield state as shown in Fig.~\ref{fig:TDCfmr}(a,b), in contrast to Fig.~\ref{fig:TDC}(a,b) where TDC-GAME and Redfield trace-distances cross much sooner.

The asymptotic GAME is also closer to the TDC-Redfield state than the Redfield state at $g_{tot}=1$.
However,  at $g_{tot}=0.01$, GAME and the Redfield states have essentially the same distance to the TDC-Redfield state. At this time it is too early to comment on this, but it suggests that the Redfield and GAME approximations are on similar, if not the same, level of approximation.
Negativity of the TDC-Redfield state is strongly suppressed at $g_{tot}=0.01$, analogous to that in Fig.~\ref{fig:TDC}(d).

More exotic initial states including the Dicke and the
Greenberger Horne Zeilinger state will be the subject of future reasearch, to establish which equation is most suitable to describe quantum correlations
and their decay. We would like to note at this time that the perpendicularly magnetized state,
which is initially a product state, evolves into a nearly maximally entangled state at approximately one period of the FMR. The study of multipartite entanglement determined by these equations is well outside the scope of the present paper, but a hint is given in Appendix~\ref{sec:floquet}.

\section{Conclusion}

The key to an effective master equation is in the combination of high accuracy and low complexity, and GAME accomplishes both with a complete positivity guarantee.
The accuracy is the result of
the combination of two steps involved in approximating the Redfield equation into a complete positive form. The first step is the renormalization of the system Hamiltonian in the Redfield equation, due to the coupling to the heat bath.
This renormalization is done before any approximation, to ensure
balance between dissipative gains and losses of the state.
The second step is to approximate the gain and loss terms to restore complete positivity.
This is accomplished by swapping out an arithmetic mean of the SD with a geometric
mean of the SD. The difference between the two means amounts to a transient term in the master equation.
The error accumulated due to its neglect saturates on time scale
of the transient, which is on the order of typical system frequency.

We test GAME on a 25-site ferromagnetic Heisenberg spin-chain with dipole-dipole magnetic anisotropy, by numerically simulating the relaxation of an initially perpendicularly magnetized state.
The density matrix is solved directly, by iterating the master equation. We compare
GAME to several other master equations, and find that GAME produces the closest state to the solution of the Redfield equation.

Over the past several years, we have studied GAME in time dependent driven open quantum systems. In fact, we derived GAME initially for time-dependent Hamiltonians, but reduced the scope of this paper to time-independent ones. The extension of GAME to time-dependent Hamiltonians is both straightforward and retains the simplicity of the (now) Floquet-Redfield equation.
In appendix~\ref{sec:floquet} we sketch the extension by presenting an example of the ferromagnetic Heisenberg spin-chain driven by a transverse time dependent magnetic field.
We study how the periodic field drives this uniaxial magnet into a long-range quantum coherent state at zero temperature.

Possible future applications of GAME include the following:

\begin{itemize}
 \item
As an extension of GAME, we plan to investigate quantum many-body systems with moderately strong coupling to the heat-bath. In such regime, $\tau_c\ll \tau_r\ll 1/E_{max}$, with $E_{max}$ the largest Bohr frequency of the system.
These studies aim to understand overdamped ground states and quantum dynamics in solid-state systems like magnets and superconducting quantum circuits. Although we expect that the Redfield equation in that regime develops states with significant negative eigenvalues and a possible instability, there is no fundamental reason not to investigate such systems within the master-equation paradigm, because the Born-Markov approximation is still valid.
A specific question is what type of CP-restoration of the Redfield theory produces a steady state that is a reasonable approximation of the reduced ground state, broadened by significant vacuum fluctuations, as well as the dynamics above that state. Possibly, this investigation could later on be extended to the true strong-coupling regime, by including the Hamiltonian renormalization and CP restoration into a larger unitary transformation such as the polaron transformation~\cite{Silbey}.

 \item Practically, GAME opens possibilities to model solid state devices
that utilize long range entanglement as resource. Complete positivity cures the headache of negative eigenvalues
 before taking the partial transpose, while low complexity of the master equation  enables studies of large number of entangled particles.
One challenge will be to determine the local-local and local-global spin correlation functions
on density matrices such as those in appendix~\ref{sec:floquet}. Quantum regression theorem is well established and can be applied in that task.~\cite{Gardiner,BreuerHeinz-Peter1961-2007TToO} These computations will be relevant for
quantum sensing of spins without an immediate spin-charge conversion, opening a way to possible applications in neutron scattering spectroscopy, high-energy physics, and gravitational astrophysics.

\item Another direction will be to apply well-established density matrix renormalization group (DMRG) methods to the study of finite and infinite spin systems with long range $1/(r^\alpha)$ interactions.
The infinite DMRG has been around for 15 or so years, and such applications have been well developed.~\cite{Gong} Tensor methods will apply well-established infinite DMRG methods to the study of this system out of equilibrium.~\cite{evenbly}

\end{itemize}

Beyond quantum-information science, it is our wish that GAME will find appeal in broader range of fields, including quantum chemistry, energy transfer dynamics, and fuel efficiency. GAME is the only quantum master equation that assures complete positivity and retains the simplicity and accuracy of the widely applied 63-year old Redfield equation.

We thank Jason Dark, Elyana Crowder, Brian Kennedy, and Glen Evenbly for discussions, and Ugo Marzolino for a comment about the discontinuity of the RWA pertinent to Sec~\ref{sec:3levsys}.
This research was supported by DOE contract DE-FG02-06ER46281, including development of the numerical simulations of quantum dynamics in magnets.
Additional support from the Georgia Tech Quantum Alliance (GTQA), a center funded by the Georgia Tech Institute of Electronics
and Nanotechnology was used to develop exact numerical methods on entangled low-dimensional magnetic systems.

\pagebreak
\section{\label{sec:App1} Appendix: Numerical Iteration of Master Equations}

We discuss how we solve a master equation
\begin{equation}
\frac{d\rho}{dt}=\mathcal{L}(\rho),
\end{equation}
within an arbitrary error ($\epsilon$) in norm distance, given the initial condition $\rho(0)$.
First we consider the case where $\mathcal{L}$ is a time independent superoperator.
The algorithm we use simulates the power series expansion of $e^{\mathcal{L}t}$, e.g., as follows.
Given a state $\rho(t)$ at time $t$, and time increment $dt$, let $\rho_d=\rho_0=\rho(t)$.
Start the loop $m=1,2,...\,$,. For each $m$,
let $\rho_d\to\mathcal{L}\{\rho_d\}$, followed by $\rho_0\to\rho_0+(dt^m)\rho_d/m!$.
Terminate the loop when $\lvert\lvert (dt^m)\rho_d/m!\rvert\rvert_2<\epsilon$.
For long simulations we use the threshold
$\epsilon=10^{-7}$, which leads to a typical error per step in the $10^{-9}$ range.
For the ferromagnetic chain we use the time increment of $dt=T_{fm}/64$.

In case of TDC master equations, we loose the above property. So we apply Runge-Kutta RK4 method in the interaction picture, and use the same time increment $dt$. (That is, a time step at time $t$ calculates the RK4-coefficients at times $t$, $t+dt/2$, and $t+dt$.)

\section{\label{sec:Invariance} Appendix: Saturation Time versus Bath Cutoff Frequency.}
\begin{wrapfigure}{L}{.7\textwidth}
\includegraphics[width=.7\textwidth]{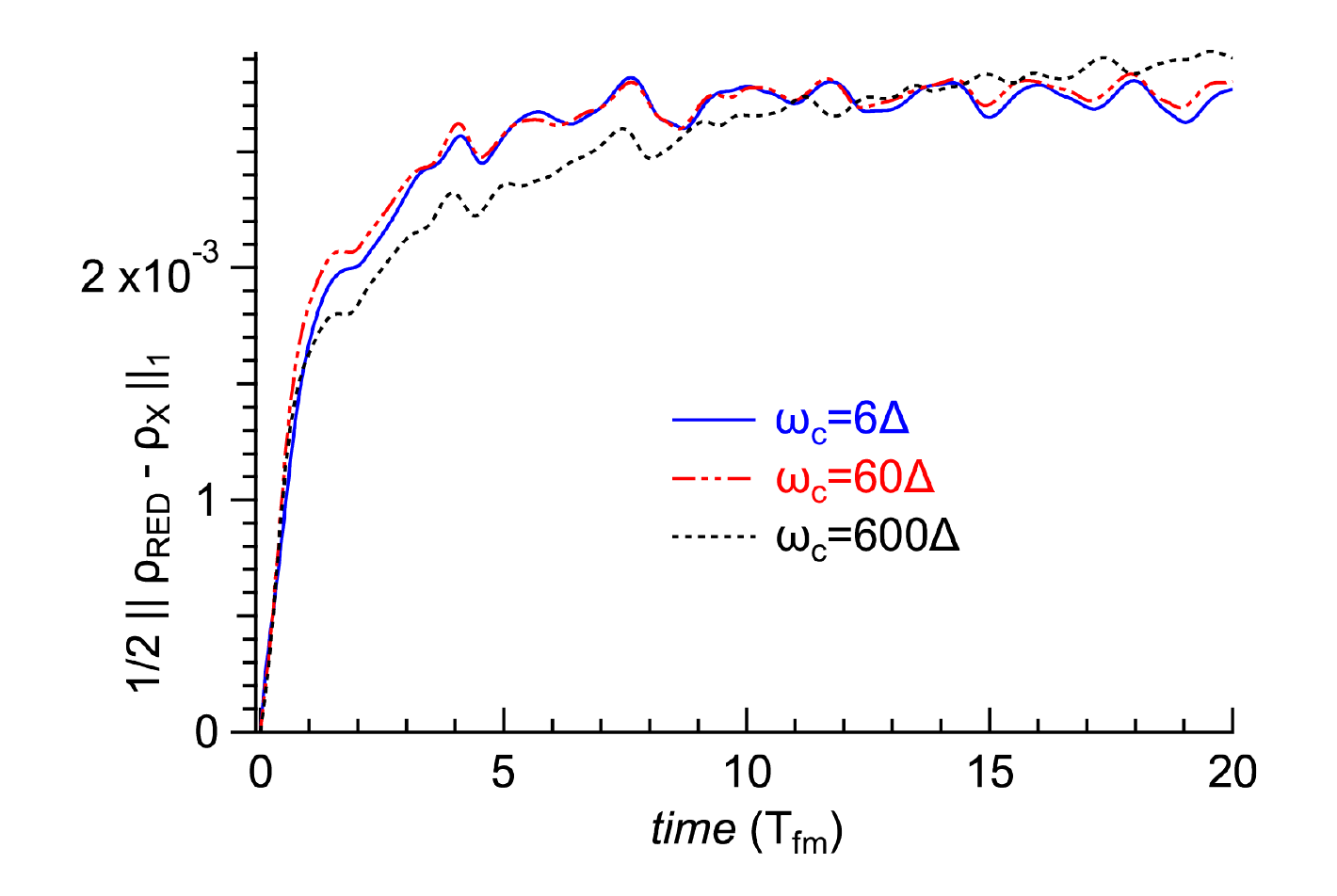}
\end{wrapfigure}
Trace distance between the solutions of the Redfield and GAME master equations versus time, at three values of $\omega_c$. $n=23$, $N=694$, $\Delta=20.1$, $T_{fm}=2\pi/\Delta$.
$g_{tot}=0.0092$, $0.0055$, and $0.0033$, at $\omega_c=6\Delta$, $60\Delta$, and $600\Delta$, respectively. The saturation time ($T_s$) of the trace-distance is weakly dependent on $\omega_c$ in this range. Fits to a single exponential (not shown) lead to $T_s=1.4$/$1.2$, at $\omega_c=6\Delta$/$60\Delta$.\\
\\
\\
\section{\label{sec:floquet} Appendix: Extension Of GAME to time-dependent Hamiltonian}

GAME can be straightforwardly extended to study the dynamics of open quantum systems with time dependent system Hamiltonian.
In that case, $H_0$ in Eq.~\ref{HamiltonianH} is replaced with time dependent periodic Hamiltonian $H_0(t)$ with period $T$.
The periodicity does not necessarily narrow the scope. For example,
if we want to study the effect of pulses, we could make the period of $H_0(t)$ much longer than other relevant time scales and rely on fast Fourier transform algorithm to rapidly
calculate the spectral functions as shown below.
We generalize GAME by rotating into the Floquet basis.

Here consider a 19-site spin-chain with the same exchange energy and anisotropy as in the main text. In this example, the dimension of the truncated Hilbert space is $N=448$,
while $H_{0}(T)=H_0-\epsilon_z S_x h(t)$, where $h(t)$ is a shape function bound by one, and $\epsilon_z$ is the Zeeman splitting at $h(t)=1$.

\begin{figure}
\centering
\includegraphics[width=.79\textwidth]{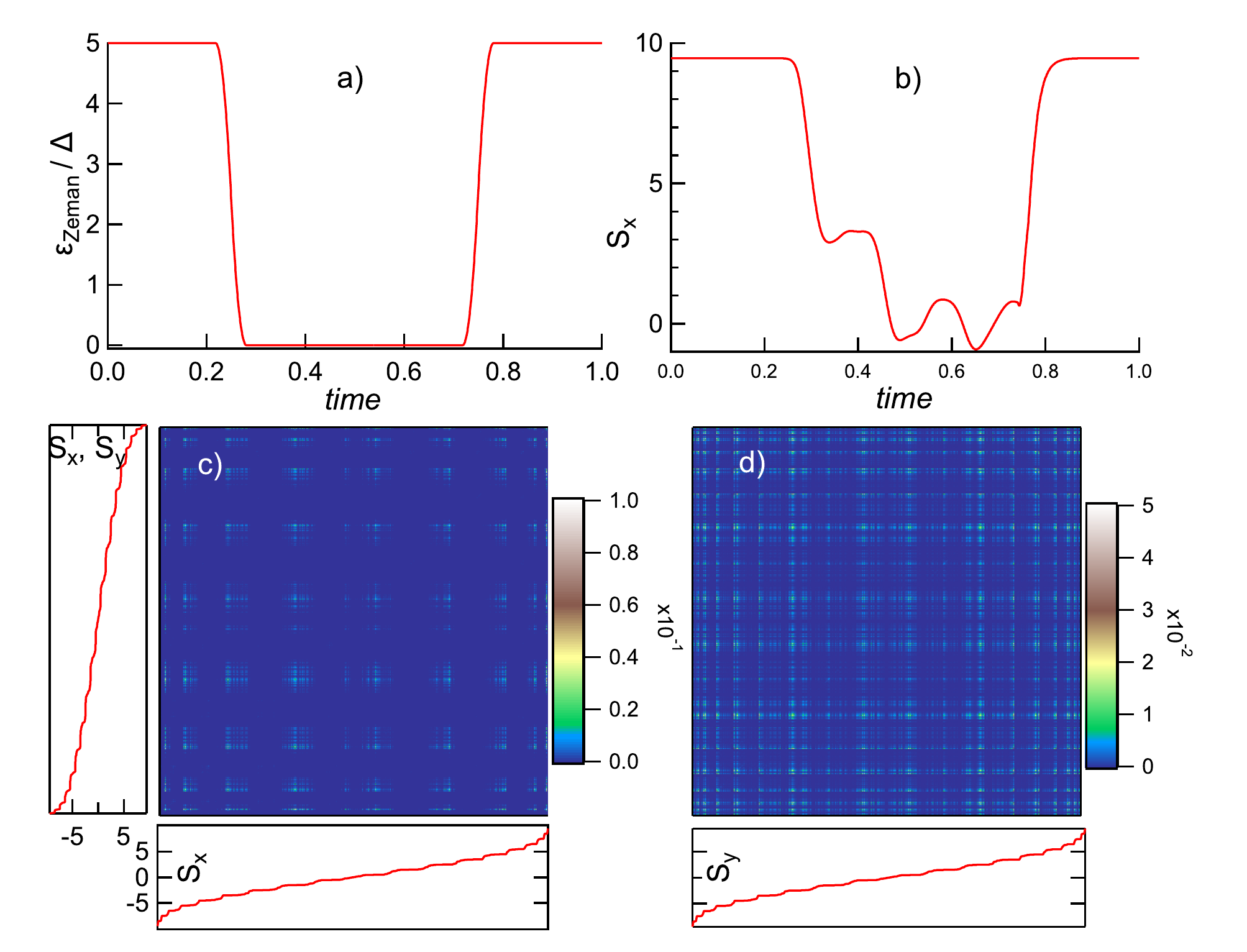}
\caption{Dynamics of a driven ferromagnetic spin-chain switches between mostly dissipative and mostly unitary during one period of the drive.
a) A nearly square field wave is applied perpendicular to the chain with period $10\times T_{fm}$. b) X-component of the magnetization versus time.
c) and d): The magnitude of the density matrix at half-period is displayed in the $S_x$ and $S_y$ basis and demonstrates coherence between oppositely magnetized states.
\label{fig:FOLQ1}}
\end{figure}

We apply a time dependent square wave of magnetic field perpendicular to the chain axis with period
$T=10T_{fm}$. In experimental setting the frequency would be roughly $1GHz$,
for typical magnets used in magneto-electronics.
Fig.~\ref{fig:FOLQ1} displays one simulated magnetic field period, with amplitude in units of the FMR gap $\Delta$. The magnetic field switches from high to zero smoothly,
is zero for approximately half the period, and then it switches back to the large value. In the high field, the ground state magnetization in the adiabatic approximation is close to aligned with the applied field.

We use a super-Ohmic spectral density given in Table~\ref{table1}, which is appropriate for
relaxation mediated by the phonon bath, with the same cutoff frequency as for the Ohmic bath ($6\Delta$) at zero temperature, and the coupling
constant $g\approx 0.2660$ per heat bath. At this $g$, the super-Ohmic spectral density leads to a similar energy relaxation time at zero field, as in the example studied in Secs.~\ref{sec:PRWA} and~\ref{sec:PERLind} at $g_{tot}=1$.

The Floquet theorem expresses the states of the system as
\begin{equation}
\vert\phi (t)\rangle=e^{-i\epsilon_\phi t}\vert u_\phi (t)\rangle,
\end{equation}
where $\epsilon_\phi$ are the quasi-energies and $\vert u_\phi (t)\rangle = \vert u_\phi (t+T)\rangle$.
The quasi-energies are determined by finding the eigenvalues $e^{-i\epsilon_\phi T}$ of the unitary time
evolution operator ($U$) for one period of the drive.
We find $U$ numerically
using RK4 with a time increment of $3.8147\times 10^{-6}T$ and verify that $\vert\vert UU^\dagger - \mathds{1}\vert\vert_1 <10^{-11}$.

Next, the density matrix and operator $A$ are represented in the Floquet basis, $\langle u_\phi\vert\rho\vert u_\xi\rangle$
and $\langle u_\phi\vert A\vert u_\xi\rangle$, respectively. As such, $A$
becomes time dependent and periodic, so we write it down as $A(t)$.
The Floquet Redfield equation~\cite{GRIFONI1998229,Hone,Shirai_2016} generalizes the Redfield Eq.~\ref{eq:redfield}:
\begin{equation}
\frac{d\rho}{dt}=-i[\text{diag}(\epsilon_\phi),\rho]-A(t)A_f^\dagger(t) \rho-\rho A_f(t)A(t)+A_f^\dagger(t)\rho A(t) +A(t)\rho A_f(t).
\label{eq:FloqRed}
\end{equation}
$A_f(t)$ is the time dependent filtered operator, which is obtained as
\begin{equation}
A_f(t)=\text{IFFT}\big\{\Gamma_m\circ\text{FFT}\big[A(t)\big]\big\}.
\label{eq:FFT}
\end{equation}
FFT and IFFT are the forward and inverse discrete Fourier transforms
in MATLAB, respectively. The frequencies we use are $(m-1)\omega_p$, $m=1, 2, ..,2048$, where $\omega_p=2\pi/T$.
$\Gamma_m$ is a matrix, related to the half ranged integral Fourier transform of the bath correlation function given by Eq.~\ref{Eq:filter}.
That is, $\Gamma_{m\phi\xi}=\Gamma(m'\omega_p+\epsilon_\phi-\epsilon_\xi)$.
FFT in MATLAB does not center frequency around zero. The easiest way to fix this issue is to shift $m$ as this:  $m'=m$, if $m\leq 1025$, and $m'=m-2049$ if $m\geq 1026$.
Virtually the entire spectrum of $A(t)$ is accounted for within our double precision calculation, because the norm of FFT$[A(t)]$ at the half frequency is approximately
$10^{-12}$ of that at zero frequency (not shown).

To establish the time-dependent version of GAME, the first step is to extract the Lamb-shift without changing the Floquet-Redfield equation. We find a time dependent renormalized Hamiltonian, by generalizing Eq.~\ref{eq:shif},
\begin{equation}
H(t)=\text{diag}(\epsilon_\Phi)-\frac{i}{2}[A(t)A_f^\dagger(t)-A_f(t)A(t)]\equiv \text{diag}(\epsilon_\Phi)+H_L(t),
\end{equation}
in the Floquet basis. The next step is to approximate the kernel (implicit in Eq.~\ref{eq:FloqRed}), and we do this by generalizing the derivation of Eq.~\ref{eq:ansatz}.
That is, we calculate a time dependent Lindblad generator
\begin{equation}
L(t)=\text{IFFT}\big\{\sqrt{\gamma_m}\circ\text{FFT}\big[A(t)\big]\big\}.
\end{equation}
 $\gamma_m$ is the SD matrix with elements $\gamma_{m\phi\xi}=\gamma(m'\omega_p+\epsilon_\phi-\epsilon_\xi)$.
In this case the transformation of the time dependent master equation into a Lindblad form is assisted by the FFT-convolution theorem.
This approximation sequence leads to a time dependent GKSL-equation,
\begin{equation}
\frac{d\rho}{dt}=-i[H(t),\rho]+\left[L(t)^\dagger\rho L(t)-\frac{1}{2}\{L(t)L(t)^\dagger,\rho\}\right].
\label{eq:GAME-ChainFL}
\end{equation}.

The magnetization x-component versus time is shown in Fig.~\ref{fig:FOLQ1}(b). When the magnetic field is high, the level
spacing of the system is high due to Zeeman splitting. The chain rapidly relaxes towards the quantum ground state, facilitated by the enhanced
super-Ohmic spectral density.
We could say that the high field part of the wave constitutes the state preparation stage.

At zero magnetic field, the level spacing is suppressed
and set by the anisotropy $\Delta$, leading to a much longer $\tau_r$. Analogous enhancement of $\tau_r$  at low field is well established in spin-1/2 quantum dots.~\cite{Elzerman,Morello}
Now we arrive to the key observation. The magnetic dynamics at zero magnetic field quickly creates a highly correlated many-body state.
At a half period, Figs.~\ref{fig:FOLQ1}(c,d) display the magnitude of the density matrix in the $Sx$ and $S_y$ eigenbasis, respectively, within the truncated Hilbert space.
The off-diagonal elements connect states with opposite
magnetization and are spread out over the entire range of $Sx$ and $Sy$. Similar coherences are found when the density matrix is represented in the $Sz$ basis (not shown). It will be interesting to calculate various correlation functions, to test if they may violate a Bell inequality in this macroscopic setting.

\section{\label{sec:TDEPSF} Appendix: Time-Dependent Spectral Functions}

The decomposition of the kernel in Eq.~\ref{eq:ddc} into single-variable functions proceeds as discussed in appendix F of Ref.~\cite{Schaller}. After the decomposition, the dissipative kernel~\ref{eq:kernelDC}
has time-dependent coefficients $\gamma_t(\omega)$ and $S_t(\omega)$ defined as
\begin{equation}
\gamma_t(\omega)=\frac{1}{\pi}\int_{-\infty}^\infty\gamma\left(\omega+\frac{x}{t}\right)\text{sinc}(x)dx
\label{eq:tdepgamma}
\end{equation}
and
\begin{equation}
S_t(\omega)=-\frac{1}{\pi}\int_{-\infty}^\infty\gamma\left(\omega+\frac{x}{t}\right)\frac{(\sin\frac{x}{2})^2}{x}dx.
\label{eq:tdepS}
\end{equation}
First we show that these are the time-dependent SD and PD introduced in Eq.~\ref{eq:GammaTDC}.

{\it Proof.} The time-dependent spectral functions $\gamma_t(\omega)$ and $S_t(\omega)$ are defined as
\begin{equation}
\Gamma_t(\omega)=\frac{1}{2}\gamma_t(\omega)+iS_t(\omega)=\int_0^t C(\tau)e^{i\omega\tau}d\tau.
\label{eq:filter2}
\end{equation}
Expressing the bath-correlation function as the inverse Fourier transform of the SD, as shown in Table~\ref{table1},
and changing the order of integrals, we find
\begin{equation}
\Gamma_t(\omega)=\frac{1}{2\pi}\int_{-\infty}^{\infty}\gamma(\Omega)d\Omega\int_0^t e^{i\left[(\omega-\Omega)\tau\right]}d\tau
=\frac{t}{2\pi}\int_{-\infty}^\infty\gamma(\Omega)e^{i\left[(\omega-\Omega)\frac{t}{2}\right]} \text{sinc}\left[(\omega-\Omega)\frac{t}{2}\right]d\Omega.
\label{eq:filter3}
\end{equation}
After substituting $\Omega=\omega+x/\tau$, this equation leads to
\begin{equation}
\frac{1}{2}\gamma_t(\omega)+iS_t(\omega)=\frac{1}{2\pi}\int_{-\infty}^{\infty}\gamma\left(\omega+\frac{x}{t}\right)
e^{-i\frac{x}{2}} \text{sinc}\frac{x}{2}\,dx.
\label{Eq:sdtransform}
\end{equation}
By taking the real and imaginary parts, we arrive at  Eqs.~\ref{eq:tdepgamma} and~\ref{eq:tdepS}, respectively. QED.

Now we demonstrate the equivalency of Eqs.~\ref{eq:dcgLAb} and Eq.~\ref{eq:dcReactance}.

{\it Proof}: Introduce the
real functions $R_t(\omega)$ and $W_t(\omega)$,
\begin{equation}
R_t(\omega)=\frac{1}{\pi}\int_{-\infty}^\infty S\left(\omega+\frac{x}{t}\right)\text{sinc}(x)dx
\label{eq:tZ}
\end{equation}
and
\begin{equation}
W_t(\omega)=-\frac{1}{\pi}\int_{-\infty}^\infty S\left(\omega+\frac{x}{t}\right)\frac{(\sin\frac{x}{2})^2}{x}dx.
\label{eq:tW}
\end{equation}

Notice that Eqs.~\ref{eq:ddc} and~\ref{eq:dcgLAb} map to each other, if we swap out the SD
with PD, and change the sign of the exponential in the prefactor. We apply this mapping to Eq.~\ref{eq:kernelDC}, by swapping $[\gamma_t(\omega),S_t(\omega)]$ with $ [R_t(\omega),W_t(\omega)]$ and changing the sign in the exponential. This leads to the unitary kernel in terms of the single variable functions of frequency $R_t(\omega)$ and $W_t(\omega)$:

\begin{equation}
\mathcal{H}_t^{dc}(\omega,\omega')=e^{i\frac{(\omega-\omega')\tau}{2}}\left\{
\frac{1}{2}[R_\tau(\omega')+R_\tau(\omega)]\text{sinc}\frac{(\omega'-\omega)\tau}{2}-
\frac{W_\tau(\omega')-W_\tau(\omega)}{(\omega'-\omega)\frac{\tau}{2}}\cos\frac{(\omega'-\omega)\tau}{2}
\label{eq:kernelDCb}
\right\}
\end{equation}

To evaluate $R_t(\omega)$ and $W_t(\omega)$, express the PD in terms of the bath correlation function (Table~\ref{table1}):
\begin{equation}
R_t(\omega)=\frac{1}{\pi}\text{Im}\int_{-\infty}^\infty \text{sinc}(x)dx\int_{0}^\infty C(\tau)e^{i(\omega\tau+\frac{\tau}{t}x)}d\tau,
\end{equation}
and
\begin{equation}
W_t(\omega)=-\frac{1}{\pi}\text{Im} \int_{-\infty}^\infty \frac{(\sin\frac{x}{2})^2}{x}dx\int_{0}^\infty C(\tau)e^{i(\omega\tau +\frac{\tau}{t}x)}d\tau.
\end{equation}
After some algebra, changing the order of integration, and applying the identity
\begin{equation}
\int_{-\infty}^{\infty} \sin (ax)\sin^2(x/2)/xdx=\frac{1}{2}\int_{-\infty}^{\infty} \cos(ax)\text{sinc}(x)dx=\Theta(1-|a|)\text{sign}(a)\pi/2,
\end{equation}
we arrive at:
\begin{equation}
R_t(\omega)=S_t(\omega),
\end{equation}
\begin{equation}
W_t(\omega)=-\frac{1}{4}\gamma_t(\omega).
\end{equation}
Substituting these into Eq.~\ref{eq:kernelDCb}, we obtain Eq.~\ref{eq:dcReactance}. QED.

For the Ohmic bath with exponential frequency cutoff, the time dependent spectral functions
can be determined explicitly,
\begin{equation}
\Gamma_t(\omega)=-ig\omega_c\left\{
1-\frac{e^{i\omega t}}{1+i\omega_ct}-\frac{\omega}{\omega_c}e^{-\frac{\omega}{\omega_c}}\left[
Ei(\frac{\omega}{\omega_c})-Ei(\frac{\omega}{\omega_c}+i\omega t)-i\pi\Theta(-\frac{\omega}{\omega_c})
\right]
\right\}.
\end{equation}
However, we find it is faster to evaluate the spectral functions by numerical integration of the
correlation function, using the function "integrate" in MATLAB. Hence, after verifying that the numerical and analytical results agree, we numerically obtain all the results related to the DCG-approximation.

\begin{figure}
\centering
\includegraphics[width=0.9\textwidth]{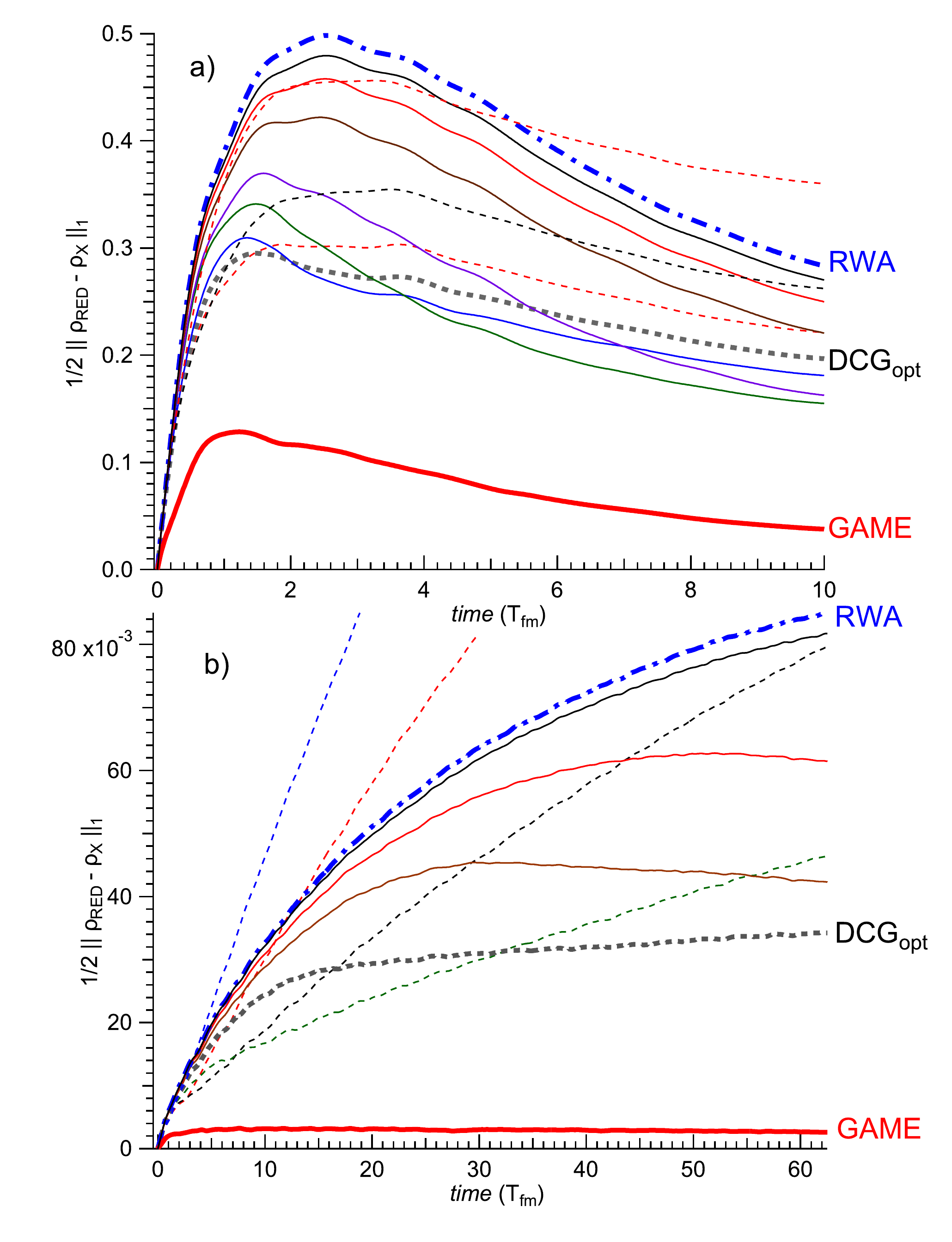}
\caption{The same as Fig.~\ref{fig:DCGA} but for Ohmic SD with the Drude-Lorentz cutoff. Trace-distances between the
solutions of various master equations and that of the Redfield equation, versus time. a) Regime of moderate environmental coupling $g_{tot}=69g=0.92$. Thick lines: RWA  (dashed-blue), $\text{DCG}_\text{opt}$ (dashed-black), and GAME (red-full). The approximate optimum coarsegraining time is $\tau_{opt}= 1.3$. Thin-full lines: coarse-graining times $\tau=64,32,16,6.4,3.2$, and $1.6$, follow the maxima between the RWA and $\text{DCG}_\text{opt}$ from top to bottom, respectively.
Thin-dashed lines: $\tau=0.96,0.64,0.32,$ and $0.16$, follow the maxima from $\text{DCG}_\text{opt}$ bottom to top, respectively.
b)  Analogous to a) in the weak environmental coupling regime $g_{tot}=0.0092$. $\tau_{opt}= 16$. Thin-full lines, top-to-bottom: $\tau=160,64,$ and $32$. Thin-dashed lines, bottom-to-top:
$\tau=6.4,3.2,1.6,$ and $0.96$.
All time scales are in units of $T_{fm}$.
$n=23$, $N=694$, $\Delta=20.1$, and $\omega_c=6\Delta$.
\label{fig:DCGB}}
\end{figure}

To check the validity of these transforms, we have solved the Redfield, GAME, and DCG master equations for Ohmic spectral density with the Drude-Lorentz cutoff, presented in table~\ref{table1}. In this calculation,
we find the time-dependent SD and PD directly from the asymptotic SD according to Eqs.~\ref{eq:tdepgamma}
and~\ref{eq:tdepS}. That way we bypass the difficult correlation function of the Drude-Lorentz SD. Similarly, we calculate the functions $Z_{t}(\omega)$ and $W_{t}(\omega)$  by numerical integration of Eqs.~\ref{eq:tZ} and Eqs.~\ref{eq:tW}. Since the Drude PD is a relatively simple function of frequency as shown in table~\ref{table1},
and contains no special functions,
these integrals are relatively fast to evaluate. That is, we do not take advantage of the explicit unitary kernel given by Eq.~\ref{eq:dcReactance}, but integrate the asymptotic coefficients over frequency. The results of the simulation of the system state versus time are displayed in Fig.~\ref{fig:DCGB}, showing a very reasonable agreement with Fig.~\ref{fig:DCGA}.

\pagebreak
\nocite{apsrev41Control}
\bibliography{master}
\end{document}